\documentclass[twocolumn,showpacs,preprintnumbers,amsmath,amssymb]{revtex4} 
 
\usepackage{graphicx}
\usepackage{dcolumn}
\usepackage{bm}
 
\newcommand{\epsfigbox}[5]{%
\begin{figure} \vspace{#3}%
\includegraphics{#2}%
\caption{ 
\label{fig:#1} #5} 
\vspace{#4} 
\end{figure}}

\newcommand{\epstblwide}[6]{%
\begin{table*}\vspace{#3}%
\caption{ 
\label{tab:#1}#5} 
\includegraphics{#2} 
\begin{flushleft}\vspace*{-4pt}#6 
\end{flushleft} 
\vspace{#4} 
\end{table*}} 
 
\begin{document}  
\title{A Multi-Shell Shell Model for Heavy Nuclei} 
 
\author{Yang Sun} 
\address{  
Department of Physics, University of Notre Dame, Notre Dame, IN 46556}   
\author{Cheng-Li Wu} 
\address{Physics Division, National Center for Theoretical Sciences,  
Hsinchu 300, Taiwan} 
 
\begin{abstract} 
Performing a shell model calculation for  
heavy nuclei has been a long-standing  
problem in nuclear physics.  
Here we propose one possible solution.  
The central idea of this proposal is to take the advantages of  
two existing models, 
the Projected Shell Model (PSM) and the Fermion Dynamical Symmetry Model (FDSM), 
to construct a multi-shell shell model.  
The PSM is an efficient method of 
coupling quasi-particle excitations  
to the high-spin 
rotational motion, whereas the FDSM contains a successful truncation  
scheme for the low-spin 
collective modes from the spherical to the well-deformed region.   
The new shell model is expected to describe simultaneously  
the single-particle and the low-lying  
collective excitations of all known types, yet  
keeping the model space tractable even for the heaviest nuclear systems. 
\end{abstract} 
 
\maketitle 
 
\section{Introduction} 
 
Except for a few nuclei lying in the vicinity of shell closures,  
most of the heavy nuclei are difficult to describe 
in a spherical shell model framework because of the unavoidable problem of 
dimension explosion. 
Therefore, the study of nuclear structure 
in heavy nuclei 
has relied mainly on the mean-field approximations, in which the concept of  
spontaneous symmetry 
breaking is applied \cite{RS80,Frau01}.  
However, there has been an increasing number of  
compelling evidences indicating that 
the nuclear many-body correlations are important.  
Thus, the necessity of a proper quantum mechanical treatment for 
nuclear states has been growing, and  
we are facing the 
challenge of understanding the nuclear structure by 
going beyond the mean-field approximations.  
 
Demand for a proper shell model treatment arises also from the nuclear 
astrophysics. Since heavy elements are made in stellar evolution and explosions, 
nuclear physics, and in particular nuclear structure far from stability, enters 
into the stellar modeling in a crucial way. 
The nucleosynthesis and the correlated energy generation 
is not completely understood, and the origin of elements in the cosmos 
remains one of the biggest unsolved physics puzzles.  
Nuclear shell models can generate 
well-defined wave functions  
in the laboratory frame, allowing us to compute,  
without further approximations as often assumed in the mean-field approaches,  
the quantities such as transition probabilities, spectroscopic factors, and 
$\beta$-decay rates. These quantities provide valuable structure information 
to nuclear astrophysics.   
In fact, the nuclear shell model calculations could strongly  
modify the results of   
nuclear astrophysics, as the recent work of Langanke and Martinez-Pinedo  
has demonstrated    
(see, for example, Ref. \cite{LM00}).  
 
Tremendous efforts have been devoted to  
extending the shell model capacity from its traditional territory 
of the $sd$-shell to  
heavier shells. Over the years, one has looked for possible solutions in the   
following two major directions. 
In the first direction, one  
employs rapidly growing computer power and sophisticated 
diagonalization algorithms to improve the traditional shell model code. The 
shell model code ANTOINE \cite{Caurier1}  
is a representative example of the recent developments  
along this line.  
Using this code the deformed $N\approx Z$ nuclei up to mass $A\sim 50$ can be  
well explained.  
The recent record example performed by this code is the full $fp$ shell 
calculation of $A=52$ nuclei \cite{ur}, with the basis dimensions in excess  
of 100 million. 
 
While in principle, it does not matter how to prepare a shell model basis, 
it is crucial in practice to use the most efficient one. 
Moreover, feasibility in computation is not our only concern. 
The other important aspect of using an efficient basis is that it 
may have a good classification scheme, such that a simple configuration 
in that basis corresponds approximately to a real mode of excitation. 
This not only can simplify the calculations, but also   
make the physical interpretations of results more easy and transparent. 
 
In the second direction, which is defined in a much wider scope, one employs   
various methods in seeking judicious truncation schemes.  
Such schemes should   
contain the most significant 
configurations, each of which  
can be a complicated combination  
in terms of the original shell model basis states.  
In this way, the basis dimension can be significantly reduced and the  
final diagonalization is performed in a much smaller space, 
thus making a shell model calculation for heavy nuclei possible.  
The early MONSTER-VAMPIRE approach \cite{MV} and the recent  
Monte Carlo 
Shell Model \cite{honma,MC} are examples along this line. 
Nevertheless, numerical calculations  
required by these models   
are still quite heavy, which may make a systematical application difficult. 
 
There are two other existing models that belong to the second category:  
the Project Shell Model (PSM) \cite{psm}, and  
the Fermion Dynamical Symmetry Model (FDSM) \cite{fdsm}. 
In the PSM, the shell model basis is constructed  
by choosing a few quasi-particle (qp) orbitals near the Fermi surfaces   
and performing angular-momentum and particle-number projection 
on the chosen configurations.  
By taking multi-qp states as the building blocks,  
the PSM has been designed to describe  
the rotational bands built 
upon qp excitations \cite{psm}.  
The PSM has been rather successful in calculating  
the high-spin states of well-deformed 
and superdeformed nuclei. 
For lighter nuclei where the large-scale shell model 
calculation is feasible \cite{Caurier2},  
studies for the deformed $^{48}$Cr \cite{HSM99}  
and the superdeformed $^{36}$Ar \cite{LS01} have demonstrated  
that the PSM calculation can  
achieve a similar accuracy in describing the data.
In the FDSM, the 
truncated basis is built by the symmetry detected $S$- and $D$-pairs, assuming 
that these are the relevant degrees of freedom for the  
low-lying collective motions. 
Having these pairs as the building blocks,  
the FDSM can provide a unified description  
for the low-spin collective excitations from the spherical to  
the well-deformed region \cite{fdsm}.  
 
It is clear that the PSM and the FDSM follow the shell model 
philosophy and both have their own shell model truncation scheme. 
However, the truncations emphasize  
different excitation modes, which are contained in one model  
but are absent in the other.  
An idea emerges naturally   
that one may combine the advantages of the two models to construct a new  
shell model for heavy nuclei.  
The question is how. 
The PSM is a microscopic approach employing the deformed intrinsic  
states and the projection  
method, while the FDSM is a fermionic model based on the group 
theory. 
The crucial step that leads us to connect these two different approaches   
is made through the recent recognition \cite{psmsu3,psmsu3b}  
that the numerical results obtained by the PSM 
exhibit, up to high angular momenta and 
excitations, a remarkable one-to-one correspondence with the analytical  
$SU(3)$ spectrum of 
the FDSM.  
This suggests that the projected deformed-BCS vacuum  
has at the microscopic level $SU(3)$-like structures which are 
very close to  
the representations of the $SU(3)$ dynamical  
symmetry of an $S$-$D$ fermion-pair system. 
This recognition has motivated us to propose a multi-shell shell model 
for heavy nuclei. Hereafter, we shall call it Heavy Shell Model,  
or HSM for short.  
  
In the following section, the PSM and the FDSM  
will be briefly reviewed. The emphasis will be given  
on the discussion of the advantages and deficiencies of each model.  
In Section 3, the connection between the two models will be explored.  
In Section 4, we will discuss in detail how  
the two models are integrated to form the HSM.  
We will give the basis states and the basis truncations for the well-deformed, 
transitional, and spherical regions. The effective interactions and the 
general method for evaluating the projected matrix elements will also be 
discussed in this section.  
Finally, the paper will be summarized in Section 5.

\section{Projected Shell Model and Fermion Dynamical Symmetry Model} 
 
In this section, we introduce 
the basic structure of the PSM and the FDSM. 
For each model, we point out the main features and the limitations. 
For interested readers, we refer to the review article of the PSM \cite{psm} 
and of the FDSM \cite{fdsm}.  
 
\subsection{Projected Shell Model} 
 
The PSM begins with the  
deformed (e.g. the  
Nilsson-type) single particle basis, with pairing correlations 
incorporated into the basis by a BCS calculation for the Nilsson states. 
The basis truncation is first implemented in the multi-qp 
states with respect to the deformed-BCS vacuum $\left|\Phi\right>$; then the 
angular-momentum and the particle-number projection are performed on the 
selected qp basis to form a shell model space in the laboratory frame; finally a 
shell model Hamiltonian is diagonalized in this projected space. 
 
If $a^\dagger_{\nu}$ and $a^\dagger_{\pi}$ are the 
qp creation operators, with the index $\nu_i$ ($\pi_i$) denoting the 
neutron (proton) quantum numbers and running over properly selected 
single-qp states, the multi-qp bases of the PSM are given as follows: 
\begin{eqnarray} 
\mbox{even-even nucleus:} && 
\{ \left|\Phi \right\rangle, \ a^\dagger_{\nu_i} a^\dagger_{\nu_j}  
\left|\Phi \right\rangle,\ 
a^\dagger_{\pi_i} a^\dagger_{\pi_j} \left|\Phi \right\rangle, 
\nonumber \\&& 
a^\dagger_{\nu_i} a^\dagger_{\nu_j} a^\dagger_{\pi_k}  
a^\dagger_{\pi_l} \left|\Phi \right\rangle, \cdots \}   
\nonumber \\ 
\mbox{odd-$\nu$ nucleus:} && 
\{ a^\dagger_{\nu_i} 
\left|\Phi \right\rangle,\ 
a^\dagger_{\nu_i} a^\dagger_{\pi_j} a^\dagger_{\pi_k} \left|\Phi  
\right\rangle,\ \cdots \}  
\\ 
\mbox{odd-$\pi$ nucleus:} && 
\{ a^\dagger_{\pi_i} 
\left|\Phi \right\rangle,\ 
a^\dagger_{\pi_i} a^\dagger_{\nu_j} a^\dagger_{\nu_k} \left|\Phi  
\right\rangle, \cdots \} \nonumber\\ 
\mbox{odd-odd nucleus:} && 
\{ a^\dagger_{\nu_i} a^\dagger_{\pi_j} \left|\Phi \right\rangle 
, \cdots \}  
\nonumber 
\label{qpset} 
\end{eqnarray} 
In bases (1), ``$\cdots$" denotes those configurations 
that contain more than two like-nucleon quasi-particles.  
If one is interested in the low-lying states only, they can practically be   
ignored because these configurations have higher excitation energies due 
to mutual blocking of levels. The bases (1) 
can be easily enlarged by including higher order 
of multi-qp states if necessary.  
If the configurations denoted by ``$\cdots$" are completely 
included, one recovers the full shell model space  
written in the representation of qp excitation.  
 
In the qp basis, the truncation can now be easily implemented  
by simply  
excluding the states with higher energies. 
Usually, only a few orbitals around the Fermi surfaces  
are sufficient for a description of the low-lying qp excitations.  
The truncation is thus 
so efficient that dimension never poses a problem  
even for superdeformed, heavy  
nuclei.  
 
After the truncation is implemented in the multi-qp  
basis, the shell model space can be constructed by the projection technique 
\cite{RS80}: 
\begin{equation} 
\left|q K IM\right\rangle=\hat{P}^{IN}_{MK}\left|\Phi_q\right\rangle 
\mbox{\hspace{6pt} with\hspace{6pt} 
$\hat{P}^{IN}_{MK}=\hat{P}^{I}_{MK}\hat{P}^{N}$,} 
\label{basis} 
\end{equation} 
where $\left|\Phi_q\right\rangle$ denotes the 
qp-basis given in (1) with $q$ meaning the multi-qp configuration, and 
\begin{eqnarray} 
\hat{P}^{I}_{MK} &=& \frac{2I+1}{8\pi^2}\int d\Omega 
D^{I}_{MK}(\Omega)\hat{R}(\Omega) 
\nonumber\\ 
\hat{P}^{N} &=& \frac{1}{2\pi}\int d\phi 
e^{-i(\hat{N}-N)\phi} 
\label{PN} 
\end{eqnarray} 
 
\noindent  
are the angular momentum and particle number projection operators, 
respectively. In Eq.\,(\ref{PN}), $D^{I}_{MK}$ is the $D$-function 
\cite{BM69}, $\hat{R}$ the rotation 
operator, $\Omega$ the solid angle, $\hat{N}$ the number operator, 
and $\phi$ the gauge angle.  
If one keeps the axial symmetry in the deformed basis,  
$D^{I}_{MK}$ 
in Eq.\,(\ref{PN}) reduces to the small $d$-function and the three  
dimensions in $\Omega$ reduce to one.  
The eigen-values $E$ and the corresponding wave functions  
$\left|\Psi^I_{M}\right\rangle =\sum_{q K}F^I_{q K}\left|q K IM\right\rangle$  
are then obtained by solving the 
following eigen-value equation:  
 
\begin{equation} 
\sum_{q K}\left\{H^I_{q K\ q' K'}-E\ N^I_{q K\ q' 
K'}\right\}\ F^I_{q' K'}=0,  
\label{sheq} 
\end{equation} 
 
\noindent  
where $H^I_{q K\ q' K'}$ and $N^I_{q K\ q' K'}$ 
are respectively the matrix elements of the Hamiltonian and the norm 
\begin{equation} 
\hspace*{-4pt} 
\begin{array}{l} 
H^I_{q K\ q' K'}\equiv\langle q K I|\hat{H}|q'K'I\rangle 
=\langle\Phi_{q}|\hat{H}\hat{P}^{IN}_{KK'}|\Phi_{q'}\rangle  
\vspace{+8pt} 
\\ 
N^I_{q K\ q' K'}\equiv\hspace{6pt}\langle q K I|q'K'I\rangle 
\hspace{6pt}=\hspace{6pt}\langle\Phi_{q}|\hat{P}^{IN}_{KK'}|\Phi_{q' 
}\rangle . 
\end{array} 
\label{Hmatrix} 
\end{equation} 
 
The PSM uses a large size of 
single-particle (s.\,p.) space,  
which ensures that the collective motion is defined 
microscopically by accommodating a sufficiently large number of active nucleons. 
It usually includes three (four) major shells each for neutrons and protons 
in a calculation for deformed (superdeformed) nuclei. 
The effective 
interactions employed in the PSM are the separable forces.   
The Hamiltonian takes 
the following form: 
\begin{equation} 
\hspace*{-4pt} 
\begin{array}{l} 
\hat H = \sum_{\sigma=\nu,\pi} \hat H_\sigma+\hat H_{\nu\pi} 
,~~~~ 
\hat H_{\nu\pi}=-\chi_{\nu\pi}\,{\hat Q^{\nu \dagger}_2} 
\hspace{-2pt}\cdot\hspace{-2pt} 
\hat Q^\pi_2  
\vspace{+8pt} 
\\ 
\hat H_\sigma = \hat H_0^\sigma 
\hspace{-2pt}-\hspace{-2pt} {\chi_\sigma\over 2} {\hat 
Q^{\sigma \dagger}_2} \hspace{-2pt}\cdot\hspace{-2pt} \hat Q^\sigma_2  
\hspace{-2pt}-\hspace{-2pt}  
G^\sigma_{\mbox{\scriptsize M}}\hat P^\sigma{}^\dagger \hat 
P^\sigma  
\hspace{-2pt}-\hspace{-2pt} G^\sigma_{\mbox {\scriptsize Q}}  {\hat 
P^\sigma_2}{}^\dagger 
\hspace{-2pt}\cdot\hspace{-2pt} 
\hat P^\sigma_2 .  
\label{H} 
\end{array} 
\end{equation} 
The first term $\hat H^\sigma_0$  in $\hat H_\sigma$ of Eq.\,(\ref{H}) is 
the spherical single-particle Hamiltonian and the remaining terms 
are residual quadrupole-quadrupole, monopole-pairing, and quadrupole-pairing 
interactions, respectively. The strength of the quadrupole-quadrupole force  
is determined 
in a self-consistent way that it would give the empirical deformation  
as predicted in the 
variation calculations. The monopole-pairing strength is taken as the 
form $G_{\mbox{\scriptsize M}} = G/A$ ($A$ is the mass number)  
with $G$ being adjusted to yield the known odd-even mass differences.  
The quadrupole-pairing strength $G_{\mbox {\scriptsize Q}}$  
is assumed to be about 20$\%$ of  
$G_{\mbox{\scriptsize M}}$ \cite{psm}. 
 
The one-body operators (for each kind of nucleons) in Eq.\,(\ref{H}) are 
of the standard form  
\begin{eqnarray} 
\hat Q_{\mu}&=&\sum_{\alpha,\alpha'}\ Q_{\mu\alpha\alpha'}\ 
c^\dagger_\alpha c_{\alpha'}  
\label{Q2m} 
\nonumber\\ 
\hat P^\dagger &=& \frac12\sum_\alpha  c^\dagger_\alpha  
c^\dagger_{\bar{\alpha}}  
\label{P}\\ 
\hat P^\dagger_{\mu}&=&  \frac12 \sum_{\alpha,\alpha'}\ 
Q_{\mu\alpha\alpha'} c^\dagger_\alpha c^\dagger_{\bar{\alpha}'} 
\label{P2m}\nonumber 
\end{eqnarray} 
where $Q_{\mu\alpha\alpha'}=\langle\alpha|\hat{Q}_{2\mu}|\alpha'\rangle$ is 
the one-body matrix element of the quadrupole operator and  
$c^\dagger_\alpha$ the nucleon creation 
operator, with $\alpha$ standing for the quantum numbers of a 
s.\,p.\ state in the spherical basis ($\alpha\equiv\{ n\ell 
jm\}$). 
The time reversal of $c^\dagger_\alpha$ is defined as  
$c_{\bar{\alpha}}\equiv (-)^{j-m}c_{n \ell j-m}$.  
 
For a description of rotational bands associated with a well-deformed minimum, 
the PSM is a highly efficient truncation scheme.  
Diagonalization for a heavy nucleus can be done almost 
instantly, yet the results are often satisfactory. The reason for the 
success is 
because the major part of pairing and quadrupole correlations has 
already been built in the basis through the use of deformed basis  
and the BCS formalism.  
Therefore, a small configuration space with a few  
s.\,p.\ orbits around the Fermi surface can already span a 
very good basis for the low-lying excitations. 
Note that each of the configurations in the basis is a complex mixture of 
multi-shell configurations of the spherical shell model space.  
Although the final dimension of the PSM  
is small, it is huge in terms of original shell model configurations. 
In this sense, the PSM is a shell model in a truncated multi-major-shell space. 
 
The pleasant features of the PSM make it a frequently-used model 
in the high-spin physics.  
Many applications can be found in the review article \cite{psm}. The recent 
publications include the study of superdeformed structure in   
a wide range of different mass regions   
\cite{Sun95,Sun97,Sun99,Ted02}, the study of the origin of identical bands 
\cite{Sun96} and of the high-K states \cite{SSW98}.   
 
While the PSM is an efficient shell model  
for deformed systems with 
rotational behavior, it becomes less valid when going to the  
transitional region, and 
eventually loses its applicability for spherical nuclei.   
Moreover, it can not efficiently describe the  
$\beta$- and $\gamma$-vibrations.  
Although such collective modes 
can in principle be obtained by mixing a  
large amount of excited qp-configurations,  
it suffers in practice from a similar dimension problem as in  
conventional shell models. 
The main reason for these shortcomings is due to 
the use of the simple BCS-vacuum which contains only the properties of the 
ground-state rotational 
band but not those of the collective vibrations.  
 
\subsection{Fermion Dynamical Symmetry Model} 
 
If we say that 
the PSM is a shell model in a truncated multi-major-shell space, then 
the FDSM is a shell model in a truncated one-major-shell space. The 
truncation is based on the consideration that the like-nucleons prefer to 
form coherent $S$ (angular-momentum $L=0$) and $D$ ($L=2$) pairs. One may thus  
assume that a closed subspace built up by these $S$-$D$ pairs is mainly 
responsible for nuclear low-lying collective motions. Such an $S$-$D$ 
subspace can be carved out by a symmetry requirement that the $S$ and $D$ 
creation (annihilation) operators together with necessary minimum 
amount of number conserving operators form a closed Lie algebra. 
To this end, a $k$-$i$ basis is introduced (see Ref. \cite{fdsm} and the 
references cited therein)  
\begin{equation} 
b^\dagger_{km_k,\, im_i}=\sum_j\  c_{km_k\, im_i}^{\ j m}\ c^\dagger_{jm} 
\label{ki} 
\end{equation} 
where $k$ (pseudo-orbital angular-momentum) and $i$ (pseudo-spin) could be 
either $k=1$ and $i =$ any half integer, or $i=3/2$ and $k =$  
any integer. This basis must uniquely reproduce the normal-parity levels   
\cite{note1}  
in that shell by $k$-$i$ angular momentum coupling,  
no more and no less. With this 
$k$-$i$ basis, the coherent $S$ and $D$ pairs and the multipole 
operators $P_{r\mu}$, which are necessary in order to form a closed Lie algebra, 
are found to be as follows 
\cite{note2}: \vspace{4pt} 
\begin{equation} 
\hspace*{-12pt}\begin{array}{c} 
S^\dagger=\sum_i\sqrt{\frac{\Omega_{ki}}2}  \left[\, b_{ki}^\dagger 
b_{ki}^\dagger\, 
\right]^{00}_{00}\\   
\mbox{(for any $k$ and $i$})\end{array},\ \ 
\Omega_{ki}=\frac{(2k+1)(2i+1)}2 
\label{S} 
\end{equation} 
\begin{equation} 
\hspace*{-10pt} 
\begin{array}{c} 
D_\mu^\dagger=\sum_i\sqrt{\frac{\Omega_{ki}}2} \left[\, 
b_{ki}^\dagger b_{ki}^\dagger\right]^{20}_{\mu0}, 
\ \ D_\mu^\dagger=\sum_i\sqrt{\frac{\Omega_{ki}}2} \left[\, 
b_{ki}^\dagger b_{ki}^\dagger\right]^{02}_{0\mu}\\ 
\mbox{($k$-active: $k=1$)} 
\hspace{64pt}\mbox{($i$-active: $i=3/2$)}  
\end{array} 
\label{D} 
\end{equation} 
\begin{equation} 
\begin{array}{c} 
\hspace{-2pt} P_{r \mu} \hspace{-2pt}=\sum_i\sqrt{2\Omega_{ki}}  \left[\, 
b_{ki}^\dagger \tilde{b}_{ki}\right]^{r0}_{\mu0},\ 
P_{r \mu}\hspace{-2pt} =\sum_i\sqrt{2\Omega_{ki}} \left[\, 
b_{ki}^\dagger \tilde{b}_{ki}\right]^{0r}_{0\mu}\\ 
\hspace{-6pt}\mbox{($k$-active: $k=1$, $r\leq 2$)} 
\hspace{20pt}\mbox{($i$-active: $i=3/2$, $r\leq 3$)}\hspace{6pt}  
\end{array}\hspace{12pt} 
\label{P2} 
\end{equation} 
\noindent  
where the symbol $[\ ]$ denotes 
angular-momentum coupling, and the time 
reversal is defined as $\tilde{b}_{km_kim_i}\equiv 
(-)^{k-m_k+i-m_i}{b}_{k-m_ki-m_i}$.  The operator set 
$\{S^\dagger, S, D^\dagger_\mu, D_\mu, P_{r\mu}\}$ forms a closed Lie algebra 
of either $Sp(6)$ ($k$-active) or $SO(8)$ ($i$-active), depending on the level 
structure of the valence shell \cite{fdsm}. 
 
Once the $S$-$D$ subspace is carved out, the form of the effective Hamiltonian 
(restricted to a two-body interaction) in 
this truncated space is uniquely determined:   
\begin{eqnarray} 
H &=& \sum_{\sigma=\nu,\pi} H^\sigma + H^{\nu\pi}\ , \hspace{14pt} 
H^{\nu\pi} = -\sum_r B_r^{\nu\pi}P^\nu_r  
\hspace{-2pt}\cdot\hspace{-2pt} P^\pi_r  
\nonumber\\ 
H^\sigma &=& H^\sigma_0 
\hspace{-2pt}-\hspace{-2pt} G^\sigma_0 
S^\sigma{}^\dagger S^\sigma 
\hspace{-2pt}-\hspace{-2pt} G^\sigma_2 
D^\sigma{}^\dagger \hspace{-2pt}\cdot\hspace{-2pt} D^\sigma  
\hspace{-2pt}-\hspace{-2pt} \sum_{r>0} B_r^\sigma 
P^\sigma_r \hspace{-2pt}\cdot\hspace{-2pt} P^\sigma_r  
\label{Hfds} 
\end{eqnarray} 
It can be shown that the multipole operators $P_r$ with $r=0,1$ are  
proportional to 
the number operator in normal parity levels $n_1$  and the total 
angular-momentum $\hat{I}$, respectively, while $P_2$ is proportional to the 
effective quadrupole operator in the truncated space.  The term 
$H^\sigma_0$ is a quadratic function of valence neutron and proton numbers 
($n^\pi n^\nu$ is included in the n-p interaction $H^{\nu\pi}$).  This  
Hamiltonian (Eq.\,(\ref{Hfds})) looks formally  
similar to that of the PSM (Eq.\,(\ref{H})) if we assume that only 
$B_2\ne 0$ in the summation.  
However, one should bear in mind that the FDSM Hamiltonian 
is written in a truncated one-major shell space and the 
s.\,p.\, energy splitting within the shell is neglected.  
This is the price the FDSM has to pay in  
order to meet the symmetry requirement so that solutions can be obtained  
with the aid of group theory. 
 
With this Hamiltonian, it can be shown that there exist analytical 
solutions in various dynamical symmetry limits, each one corresponding to a 
collective mode known experimentally: 
the $SU(3)$-limit in the $k$-active shell and the $SO(6)$-limit in the  
$i$-active 
shell correspond respectively   
to rigid and $\gamma$-soft rotors for well deformed nuclei, 
while the $SU(2)$-limit in the $k$-active shell and the $SO(5)$-limit in the 
$i$-active shell correspond to a vibrator of spherical nuclei.  
Since the FDSM contains all major collective modes, the general feature  
of different collective motions arises naturally  
as the number of valence nucleons varies;  
namely, nuclei behave as a spherical vibrator near the closed shell and  
become a well-deformed rotor around the 
mid-shell.  
If the strengths of the Hamiltonian are properly chosen and 
the s.\,p.\, splitting is taken into account as a perturbation,  
the FDSM can even quantitatively reproduce the low-lying spectra,  
B(E2)'s, ground state masses, etc., in a unified manner  
from the spherical to well-deformed region.  
 
\epstblwide{tb 1}{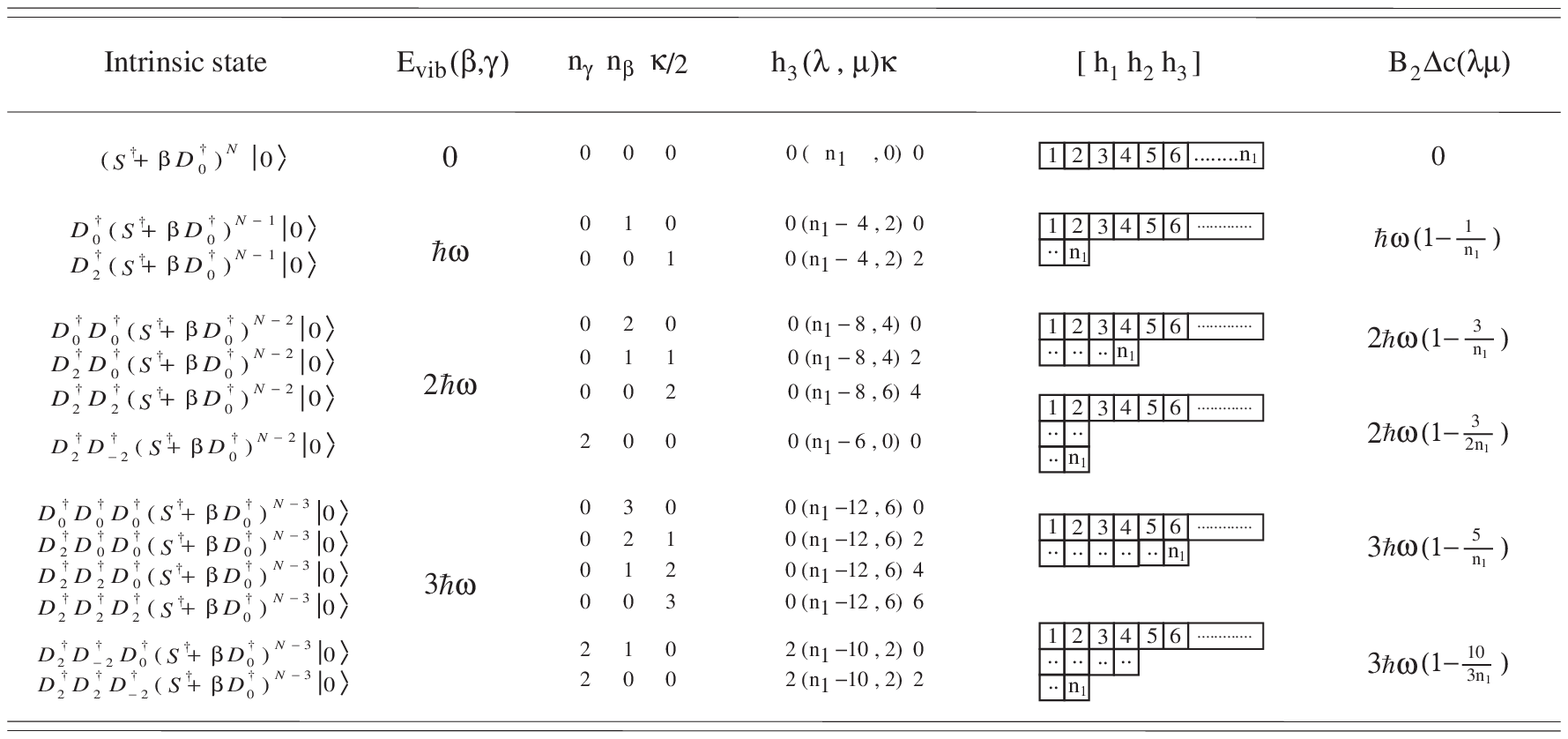}{0pt}{0pt} 
{The $SU(3)$ irreps in the 
FDSM and the collective  
$\beta$-$\gamma$ vibrations. *} {* The $SU(3)$ intrinsic 
states for each of the  
$\beta$-$\gamma$ vibrational modes are listed in the first column. The second  
and 
the third column are the phonon excitation energies and the associated quantum 
numbers according to the Particle-Rotor Model.  
The 4th and 5th column are the $SU(3)$ irreps 
$(\lambda,\mu)$ and the corresponding Young tablets  
$[\mbox{h}_1,\mbox{h}_2,\mbox{h}_3]$, respectively. The excitation 
energies obtained from the FDSM are listed in the last column as  
$B_2\Delta C(\lambda,\mu)$, where $\Delta C(\lambda,\mu)$ is the  
changes of the expectation values  
of the $SU(3)$ Casimir operator with respect  
to the ground state irrep $C(n_1,0)$, $B_2$ is the 
$n$-$p$ quadrupole interaction strength, and $\hbar\omega\equiv  
(3/2)\,n_1B_2$.} 
 
Here, it may be appropriate to emphasize one remarkable result of the FDSM:  
There exists an one-to-one correspondence between the   
$SU(3)$ irreps and the $\beta$- and $\gamma$-vibrations in deformed nuclei.   
As one can see from Table I, the $\beta$-$\gamma$ vibrations are  
microscopically 
the collective $D$-pair excitations (the $D^\dagger_{\pm 1}$ excitations are 
forbidden by the time reversal symmetry). The anharmonic behavior of the  
$\beta$-$\gamma$ vibrations is due to the finite particle number effect. In the 
large $n_1$ limit, ignoring $1/n_1$, the FDSM reproduces exactly the   
Particle-Rotor Model results.  
This means that the FDSM has discovered   
the relevant 
fermion degrees of freedom of nuclear collective motions, which are the $S$ and 
$D$ pairs.  
The $S$-$D$ subspace  
is so compact that it never suffers from the dimension explosion even for the 
heaviest nuclear systems.  
 
While with the $S$-$D$ subspace the FDSM is able to provide 
a microscopic view to the low-lying collective motions, 
it has difficulties in describing the   
qp excitations and the high-spin physics for lack of the s.\,p.\ degrees of 
freedom.  In principle, this can be resolved by allowing  
a few pairs to break.  
The problem is that once the s.\,p.\ degrees of freedom open up, the dimension 
increases very rapidly.  In addition,  
inclusion 
of s.\,p.\ degrees of freedom results in adding many new terms to the  
effective 
Hamiltonian. To pin down so many coupling strengths in the Hamiltonian is a 
very difficult task, if not impossible.  Therefore, even the number of broken  
pairs is limited to just one for proton and one for neutron,  
the model could at best be   
applied to the $fp$ shell nuclei, and could not go any further.

\section{Connection between PSM and FDSM} 
 
Let us summarize our main claims in the last section. 
Both the PSM and the FDSM are truncated shell model,  
aiming at grasping the essential ingredients to describe  
the low-lying physics.  
However, the emphasis in each model is different, which is reflected in their  
different truncation schemes. The PSM emphasizes the high-spin 
description of rotational states built upon qp-excitations  
associated with a well-deformed 
minimum, but it is not an efficient method for the collective vibrations. 
In contrast, the FDSM has a well-defined classification  
for all known types of 
collective vibration, ranging from the spherical, via transitional, 
to the well-deformed region, but it is lack of the necessary degrees of freedom 
of qp-excitations. Thus,  
the main advantages of the PSM and the FDSM are 
mutually complementary to each other.  
 
If we could take the advantages of the PSM and the FDSM and combine them 
into a single model, the  
deficiencies of each model will be eliminated.  
At first glance, it is not obvious  
how to bridge these two different approaches.  
In this section, we show that a realization  
of the combination is possible.  
The assertion is made based on the recent recognition that  
an $SU(3)$  
symmetry can emerge from the projected deformed-BCS vacuum. 
Having this as the basis, the full idea of the microscopic   
classification for the collective vibrations discovered by the 
FDSM could be adopted by the PSM while the latter keeps its original features 
in building the shell model configuration space through projection.   
 
\subsection{Emergence of $SU(3)$ Symmetry in PSM}  
 
\epsfigbox{fg 1}{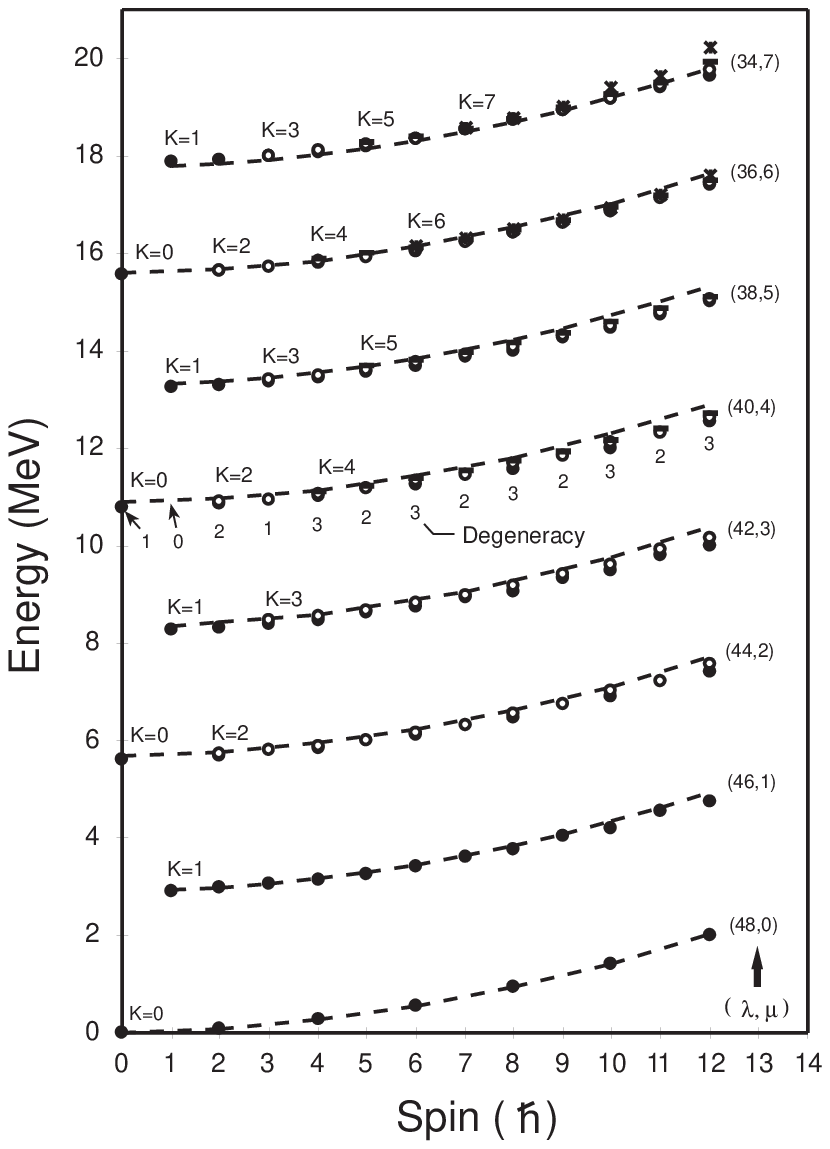}{0pt}{0 pt}{ 
Comparison between the PSM (symbols) 
and the FDSM $SU(3)$ limits (curves) assuming $n^{\mbox{eff}}_{\nu}=32$ 
and $n^{\mbox{eff}}_\pi=16$.  The $SU(3)$ quantum numbers 
$(\lambda,\mu)\kappa$ are marked for each band with degeneracy of 
$\kappa=\kappa_{max},\kappa_{max}-2,\cdots, 0 
\mbox{ or }1$, and $\kappa_{max}=min\{\lambda,\mu\}$.  
The comparison takes the well-deformed nucleus $^{168}$Er 
as an example.} 
 
It is remarkable \cite{psmsu3,psmsu3b} 
that the states numerically obtained by the PSM 
exhibit a beautiful one-to-one correspondence with the analytical  
$SU(3)$ spectrum of the FDSM.  
To show this, the original PSM 
was extended in such a way that instead of a single BCS vacuum, the angular 
momentum projection was performed for separate neutron and proton BCS vacuum 
with the same deformation, and the projected states were  
coupled through diagonalization of the 
Hamiltonian \cite{psmsu3}.  
The extension is necessary; Otherwise the PSM would have no  
collectively excited bands to compare with the FDSM. 
This procedure gives the usual rotational ground-band 
corresponding to a strongly coupled BCS condensate of neutrons and protons, but 
also led to a new set of excited bands arising from the vibrations of relative 
orientation of the neutron and proton cores (the so-called ``scissors" mode).  
  
As one can see from Fig.\,1, the spectrum obtained from the PSM is,  
up to high angular momenta and excitation energies,  
nearly identical  
to that from the FDSM $SU(3)$ formulism with $(n^{\mbox{\scriptsize 
eff}}_\nu=32, 0)$ and 
$(n^{\mbox{\scriptsize eff}}_\pi=16, 0)$ irreps,  
respectively assigned to the neutron and 
proton BCS vacuum. The classification of the spectrum 
follows exactly the FDSM $SU(3)$  
reduction rules 
\begin{equation} 
(n^{\mbox{\scriptsize eff}}_\nu, 0)\otimes(n^{\mbox{\scriptsize 
eff}}_\pi, 0)\supset (\lambda,\mu)\ \kappa I  
\end{equation} 
with 
\begin{eqnarray} 
\lambda&=&n^{\mbox{\scriptsize eff}}_\nu+n^{\mbox{\scriptsize 
eff}}_\pi-2\mu 
\nonumber 
\vspace{4pt}\\  
\mu&=&\mu_{\mbox{\scriptsize max}}, \mu_{\mbox{\scriptsize max}}-1,\cdots,1,0; 
\ \ \ \mu_{\mbox{\scriptsize max}}={\mbox{min}}(n^{\mbox{\scriptsize eff}}_\nu,  
                                                n^{\mbox{\scriptsize eff}}_\pi) 
\nonumber 
\end{eqnarray} 
and 
\begin{eqnarray} 
\kappa&=&\kappa_{\mbox{\scriptsize max}}, \kappa_{\mbox{\scriptsize max}}-2, 
\cdots, 1\mbox{ or }0;  
\ \ \ \kappa_{\mbox{\scriptsize max}}={\mbox{min}} (\lambda,\mu) 
\nonumber 
\vspace{4pt}\\ 
I&=&\left\{ 
\begin{array}{l} 
\lambda+\mu,\lambda+\mu-2, \cdots, 1 \mbox{ or } 0\ ,\ \ 
\mbox{if }\kappa=0\vspace{4pt}\\ 
\kappa,\ \kappa+1, \cdots, 
\lambda+\mu-\kappa+1,\ \ \mbox{if }\kappa\ne0 . 
\end{array}\right. 
\nonumber 
\end{eqnarray} 
 
\noindent  
Not only the spectrum but also the B(E2)'s exhibit this correspondence  
\cite{psmsu3b}.  
Note that there are many types of $SU(3)$ models, and  
that different $SU(3)$ models have different  
permissible irreps and reduction rules  
due to the different physical input,  
and therefore can lead to different band structures. 
Here, we emphasize that this 
$SU(3)$ symmetry shown in Fig. 1 is of the FDSM type.  This is because  
among the existing fermionic $SU(3)$ models in nuclear physics, only the FDSM 
$SU(3)$ formulism can naturally provide the required irreps and reproduce the 
band structure of the PSM. Other $SU(3)$ models such as 
the pseudo-$SU(3)$ do not have this property. 
In their recent investigation of the onset of rotational motion, Zuker {\it et. al.} 
\cite{qsu3a,qsu3b} 
introduced a preliminary formulation of an approximate quasi-$SU(3)$ symmetry.
It would be interesting to study whether the quasi-$SU(3)$ 
contains a similar property. 
  
Fig.1 presents a highly non-trivial result  
because the PSM, as described above,   
is not built on any explicit $SU(3)$ symmetry, and no free  
parameters have been adjusted to obtain such a symmetry.  
Nevertheless, the spectra, the electromagnetic transition rates, and the  
wave functions of the PSM agree nearly perfectly  
with the FDSM $SU(3)$ results, from the ground-band to states of 
high spins and  high excitations \cite{psmsu3b}.  
This strongly suggests that the projected  
deformed-BCS vacuum in the PSM 
could be at the microscopic level  
close to the $S$-$D$ core in the FDSM, provided that the 
$S$-$D$ pairs must be redefined in a multi-major-shell space, since the  
one-major-shell is not large enough to accommodate so many active nucleons 
(here $n^{\mbox{\scriptsize eff}}_{\nu}=32$  
and $n^{\mbox{\scriptsize eff}}_\pi=16$). 
 
In this regard, there is a 
conceptual distinction between the 
$SU(3)$ symmetry in the one-major-shell FDSM and that emerged from the 
PSM.  In the former, the $SU(3)$ symmetry arises entirely from the 
normal-parity nucleons, and the abnormal parity orbit enters only implicitly 
through the Pauli effect and the renormalization of the parameters.  In 
the latter, the $SU(3)$ symmetry arises from the explicit dynamical 
participation of both normal and abnormal-parity nucleons of many shells.   
Ignoring the direct  
contribution from the abnormal-parity nucleons in the one-major-shell FDSM is a 
sacrifice for having an exact symmetry (which forms a closed Lie algebra).  
In practice, this turns out to be not a bad approximation for collective 
motions because in a single 
major shell, there is only one single-$j$ level with abnormal-parity,  
which does not have much quadrupole collectivity compared to the normal-parity 
contributions \cite{abnormal}.  
When going to a multi-major-shell space, the 
situation will change.  A bunch of abnormal-parity levels, which are located 
just below the normal-parity levels, will open up.  
This means that, when one redefines the FDSM-type coherent $S$-$D$ pairs in a 
multi-major-shell space, the conceptual distinction of the $SU(3)$ symmetry 
between the FDSM and the PSM will be eliminated. 
 
\subsection{Collective D-Pair Excitations} 
 
To see further the relationship between the two models, let us  
ignore the terms with $r\ne 2$  
in the most general FDSM Hamiltonian (\ref{Hfds}). This is a  
reasonable approach because   
from the multipole expansion point of view, there is no multipole interactions  
with $r=odd$ without considering the parity admixture.   
One may notice that there is the monopole-monopole  
interaction ($r=0$) in the FDSM, which is not included in the  
PSM. However, it is 
well-known that this interaction only affects the nuclear total binding energy  
but does not have  
much influence on the excitations.   
On the other hand, it can be added to the PSM if necessary.  
Thus, we have for both models the interactions of  
the monopole- and quadrupole-pairing plus the 
quadrupole-quadrupole type.   
 
The PSM Hamiltonian contains operators written in the 
ordinary shell model basis, whereas the FDSM ones in the $k$-$i$ basis. 
In order to compare them, we transform the FDSM operators  
in Eq.\,(\ref{S}--\ref{P2}) back to the ordinary shell model basis  
by rewriting them in terms of   
$c^\dagger_\alpha$ and $c_\alpha$.  
We find that 
\begin{eqnarray} 
P_{2\mu} &=& \sum_{\alpha,\alpha'}\ Q^{(2)}_{\mu\alpha\alpha'}\ 
c^\dagger_\alpha c_{\alpha'} \vspace{2pt} 
\nonumber 
\\ 
S^\dagger &=& \frac12\sum_\alpha  c^\dagger_\alpha  
c^\dagger_{\bar{\alpha}}\vspace{2pt}\\ 
\label{QPfds} 
D^\dagger_\mu &=& \frac12 \sum_{\alpha,\alpha'}\ 
Q^{(2)}_{\mu\alpha\alpha'} c^\dagger_\alpha c^\dagger_{\bar{\alpha}'} 
\nonumber 
\end{eqnarray} 
with \vspace{-8pt} 
\begin{eqnarray} 
Q^{(2)}_{\mu\alpha\alpha'}&=&\sqrt{2\Omega_{ki}}C^{\ 2 \mu}_{jm\ j'-m'} 
(-)^{r-\mu}\left[ 
\begin{array}{lll} 
i&k&j\\ 
i&k&j'\\ 
0&2&2 
\end{array}\right] . 
\nonumber 
\end{eqnarray} 
Comparing Eq.\,(14) with Eq.\,(\ref{P}), we see that the 
definitions of 
$S^\dagger$, $D_\mu^\dagger$ and $P_{2\mu}$ in the FDSM are similar to 
that of $\hat{P}^\dagger$, $\hat{P}^\dagger_\mu$ and $\hat{Q}_\mu$ in 
the PSM, except that  
the model spaces for the two models are different.  However, within  
one-major-shell,  
the operator $S^\dagger$ ($S$) is exactly the same as $P^\dagger$ 
($P$) in the PSM.  For the $D$-pair and $P_{2}$ operators,   
although the coefficients $Q^{(2)}_{\mu\alpha\alpha'}$ look different from  
$Q_{\mu\alpha\alpha'}$ appearing in  
the corresponding operators $\hat{P}^\dagger_\mu$ and $\hat{Q}_\mu$ 
in the PSM, their physical meanings are the same. The FDSM Hamiltonian may  
thus be considered as an one-major-shell version of the PSM Hamiltonian, with the 
approximations of  $\hat{P}^\dagger_\mu$ 
and $\hat{Q}_\mu$ being replaced by $D_\mu^\dagger$ and $P_{2\mu}$ and the s.\,p.\ 
energy splitting being ignored.   
In other words, if, in the
multi-major-shell case, the symmetry constraint is released from the 
FDSM and the s.\,p.\
energy splitting is considered,
$S^\dagger$, 
$D_\mu^\dagger$ and $P_{2\mu}$ in the FDSM should return back to the version of  
$\hat{P}^\dagger$, $\hat{P}^\dagger_\mu$ and $\hat{Q}_\mu$ operators in the PSM. 
So is for the Hamiltonian.  
 
From the above analysis we may  
conclude that the PSM and the FDSM are just two 
approaches to solve an effective Hamiltonian of a common form.   
In order for them to be 
applicable to heavy nuclei, approximations have to be made 
in each model.  The PSM is 
based on a multi-major-shell space so that it can describe the nuclear 
rotational motion microscopically  
through a dynamic participation of many particles.   
However, it can afford to do so only for a truncated configuration 
space that includes only a BCS vacuum  
plus a few qp excitations. As a sacrifice, this 
truncation do not include the collective modes such as  
$\beta$- and $\gamma$-vibrations.  In 
contrast, the FDSM aims at nuclear low-lying collective excitations.  It can  
afford to do so only when the model space is reduced to one-major shell like in 
the  conventional shell model. But that is still not enough for heavy nuclei.  
Additional approximations to further reduce the configuration  
space down to the symmetry detected $S$-$D$ subspace are necessary.  These 
approximations exclude the s.\,p.\ excitations.

\section{Construction of Heavy Shell Model} 
 
Having realized that 
the projected deformed-BCS states exhibit the $SU(3)$ symmetry,  
and that the collective excitations may be approached by the $D$-pair excitations, 
with the $D$-pair operator defined as the quadrupole-pair operator 
$\hat{P}^\dagger_\mu$  
in a multi-major-shell model space,  
we propose a multi-shell shell model: the Heavy Shell Model. The 
essence of this proposal  
is to adopt the truncation scheme for the collective modes, which was   
discovered by the FDSM, into the PSM to enrich the shell model basis.  This 
essentially combines the advantages of both models.  

In fact, 
to incorporate both  s.\,p.\  and collective excitations, there are in principle two 
alternatives: One can either extend the FDSM by adding the PSM qp truncation 
scheme on top of the FDSM collective states, 
or extend the PSM by including the $D$-pair collective 
excitations into the PSM vacuum.  However, the FDSM is 
a severely truncated one-major-shell shell model dictated by symmetry, 
and in this sense, it is not as microscopic as the PSM. 
Although it is quite successful in the description of low-lying 
collective motions, the FDSM is just an effective 
theory. For practical applications, large renormalization effects must be embedded in the 
parameters of the FDSM Hamiltonian, which have to be determined phenomenologically.   
Therefore, we choose to construct the Heavy Shell Model based on the extension of 
the PSM.
 
The main 
ingredients of the Heavy Shell Model are as follows: 
 
\begin{enumerate} 
 
\item[1.]  Keep the multi-major-shell basis 
of the PSM as the model space, using the PSM 
$P^\dagger$ and $P_2^\dagger$ operators to describe the coherent 
$S$- and $D$-pairs in the multi-major-shell configurations, and construct the 
intrinsic collective excitation states by 
$P_2^\dagger$ acting on the deformed-BCS vacuum; 
\item[2.] Carry out the shell model truncation by selecting a few 
single-qp states near the Fermi surfaces plus a few $D$-pair excitations, and  
perform angular momentum and particle number  
projection to obtain a shell-model basis in the laboratory frame; 
\item[3.] Keep the PSM Hamiltonian to be the effective Hamiltonian, but allow 
adding more multipole interactions and/or readjusting the interaction strengths 
if necessary; 
\item[4.] Utilize the algorithms developed in the PSM  
to carry out calculations for all the necessary matrix elements, 
and diagonalize the 
Hamiltonian in the truncated shell model space. 
\end{enumerate} 
 
Let us now discuss each of the items in more detail. 
 
\subsection{The Basis States} 
 
We have 
demonstrated that the projected  
deformed-BCS vacuum $|\Phi\rangle$ in the PSM is nearly   
identical to the FDSM-$SU(3)$ intrinsic ground state  
($n^{\mbox{\scriptsize eff}},0$) irrep. 
Furthermore, we have indicated that the FDSM $S$- and 
$D$-pair and the $P_{2\mu}$ operator  
are, respectively, the symmetry-constraint 
one-major-shell version of the ordinary monopole pair $P$,  
quadrupole pair 
$P_\mu$, and quadrupole operator $\hat{Q}_\mu$ in the PSM. 
It is therefore natural to believe that in a multi-major-shell space, the  
FDSM $S$- and $D$-pairs are nothing but the $P$- and $P_\mu$-pairs   
if we abandon the symmetry requirement. 
So is for the 
quadrupole operator. In the FDSM, all known types of low-lying  
collective excitation  
can be obtained by acting   
$D^\dagger_\mu$ on the FDSM-$SU(3)$ intrinsic ground state, 
$(S^\dagger+\beta D_0^\dagger)^N|0\rangle$ (see Table I). Combining 
these facts, the collective excitations of the HSM  
in a multi-major-shell space may be 
constructed by replacing, respectively, 
$(S^\dagger+\beta D_0^\dagger)^N|0\rangle$ and 
$D^\dagger$ in Table I with  $|\Phi\rangle$  and $P^\dagger_{\mu}$  
defined in the PSM.  
 
Hereafter, we will continue to use the FDSM notations $S$ and $D$ 
for the pair operators. One should bear in mind  
that they have been redefined as  
\begin{eqnarray} 
S^\dagger &\equiv&  
P^\dagger=\frac12\sum_{\alpha}\ c^\dagger_\alpha 
c^\dagger_{\bar{\alpha}} 
\nonumber\\ 
D^\dagger_{\mu} &\equiv& \hat 
P^\dagger_{\mu}=\sum_{\alpha,\alpha'}\ 
Q_{\mu\alpha\alpha'}\ c^\dagger_\alpha c^\dagger_{\alpha'} .  
\label{SDext} 
\end{eqnarray} 
The intrinsic collective states can then be expressed as 
\begin{eqnarray} 
|\Phi_{c}\rangle 
&\equiv& \left|N_D (n_\beta,n_\gamma,\kappa)\right\rangle 
=\prod_{i=0}^{N_D} 
D^\dagger_{\mu_i}|\Phi\rangle\vspace{2pt} 
\nonumber\\ 
&=& (D^\dagger_0)^{n_\beta} 
(D^\dagger_2D^\dagger_{-2})^{\frac{\mbox{\scriptsize$n$}_\gamma}{2}} 
(D^\dagger_2)^{\frac{\mbox{\scriptsize$\kappa$}}{2}}\left|\Phi\right\rangle .  
\label{sdcore} 
\end{eqnarray} 
Eq. (\ref{sdcore}) provides the microscopic meaning of the quantum 
numbers $n_\beta$, $n_\gamma$ and $\kappa$ in a very clear manner: the phonon  
number appearing in phenomenological models is nothing  
but the total number of 
$D$-pairs 
\begin{equation} 
N_D=n_\beta+n_\gamma+\kappa/2. 
\label{nd} 
\end{equation} 
 
The basis of the HSM can be constructed by adding qp-excitations on  
top of the collective intrinsic states; the formalism is the same as 
that used to build the PSM bases in   
Eq.\,(1), but the simple  
BCS vacuum $|\Phi\rangle$ in Eq.\,(1) is now replaced by  
a more correlated one, $|\Phi_c\rangle$.  The general expression of 
the HSM basis in the laboratory system can be written as  
\begin{eqnarray} 
\left|qc IM\right \rangle &=& 
\hat{P}^{IN}_{M}\left|\Phi_{qc}\right\rangle\vspace{4pt} 
\nonumber\\ 
\left|\Phi_{qc}\right\rangle 
&\equiv& \prod_{i=0}^{n^\nu_{q}}\prod_{j=0}^{n^\pi_{q}}a^\dagger_{\nu_i} 
a^\dagger_{\pi_j}\left|\Phi_{c}\right\rangle ,  
\label{newbasis} 
\end{eqnarray} 
where $n^\nu_{q}$ ($n^\pi_{q}$) is  
the qp number of neutrons (protons), 
and the indices $q$ and $c$ stand for the qp and the collective 
vibrational configurations, respectively.   
 
The HSM basis (\ref{newbasis}) contains both s.\,p.\,   
(qp excitations) and collective 
($D$-pair excitations) degrees of freedom;  
the deficiency of lack of collective degrees of freedom in the original 
PSM is redeemed.  
Moreover, the basis (\ref{newbasis}) is expected to work  
also for the transitional (or  
weakly deformed) nuclei, which is beyond the original PSM territory.  
This is expected because  
it is known from the FDSM that the collective states of transitional 
nuclei can be described as a mixture of different $SU(3)$ irreps \cite{fdsm}. 
The original PSM 
uses only the deformed-BCS vacuum  
(the ground state of the $SU(3)$ irreps), and thus 
does not contain such a mixing mechanism.   
This is why the PSM becomes less and less 
valid when going  
away from the well-deformed region.  The HSM basis (\ref{newbasis}) 
now contains all 
possible $SU(3)$ irreps, since its labels ($ n_\beta$,  
$n_\gamma$, $\kappa$) are in one-to-one correspondence to that of the $SU(3)$  
irreps $(\lambda,\mu)$ \cite{bgvib}.  
 
For spherical nuclei, the basis should be 
constructed separately since in the spherical case, the rotational symmetry is 
restored so that  
no distinction can be made between the intrinsic and the laboratory 
system.  In the spherical limit, the BCS vacuum is just the $I=0$ ground state,  
which, in the FDSM, corresponds to the intrinsic state with  
$\beta\rightarrow 0$ (see Table II). Let us denote this BCS vacuum as 
$|\Phi_0\rangle$ to distinguish it from the deformed one. The deformed 
mean-field ({\em e.\,g.\,} the Nilsson scheme) in the spherical 
case is also reduced to  
the spherical shell model level scheme so that a qp state can be labeled by 
\{$N\ell jm$\}.  Using \{$jm$\} for short, a qp state can be expressed as 
$a^{\sigma}_{j}{}^\dagger_{m}|\Phi_0\rangle$ ($\sigma=\nu,\pi$), which is  
no longer a superposition of all possible angular momentum 
states but has a definite angular momentum. Likewise, a $D$-pair excited  
state $D^\dagger_{\mu}\left|\Phi_0\right\rangle$ is now a $2^+$ 
state. Therefore, there is no need for performing  
angular-momentum projection.  The 
basis with a given total angular-momentum $I$ can be directly obtained by the   
angular-momentum coupling.  The general expression of the HSM basis  
in the spherical case  
can be written as follows: 
 
\begin{eqnarray} 
\left|qcIM\right\rangle&=&\hat{P}^N\left|\Phi^{IM}_{qc}\right\rangle 
\vspace{2pt}\nonumber\\ 
\left|\Phi^{IM}_{qc}\right\rangle 
&\equiv&  \left[{\bf A}^\dagger_{J_q}(n_{q})\otimes 
{\bf D}^\dagger_{R_c}(N_D)\right]^I_M\left|\Phi_0\right\rangle  
\label{sbasis} 
\end{eqnarray} 
with 
\begin{equation} 
{\bf A}^\dagger_{J_qM_q}(n_{q})\hspace{7pt}\equiv   
\left[A^\dagger_{J_q^\nu}(n^\nu_{q})\otimes 
A^\dagger_{J_q^\pi}(n^\pi_q)\right]^{J_q}_{M_q} 
\label{aq}\vspace{-12pt} 
\end{equation} 
 
\begin{eqnarray} 
A^{\dagger}_{J_q^\sigma M_q^\sigma}(n^\sigma_q) 
\hspace{-2pt}\equiv\hspace{-2pt}\mbox{\Large$[$}\prod_{i=1}^{n^\sigma_q} 
a^\dagger_{j_i} 
\mbox{\Large$]$}^{J^\sigma_q}_{M^\sigma_q} 
\hspace{-4pt}=\hspace{-4pt}\sum_{[m]}C_{[m]}({J_{q}^\sigma M_q^\sigma}) 
\prod_{i=1}^{n^\sigma_q}a^\dagger_{j_i m_i}\hspace{4pt}&&  
\label{aqm}\\ 
{\bf D}^{\dagger}_{R_cM_c}(N_D)\hspace{-2pt}\equiv\hspace{-2pt} 
\mbox{\Large$[$}\prod_{k=1}^{N_D}D^\dagger\mbox{\Large$]$}^{R_c}_{M_c} 
\hspace{-2pt}=\hspace{-2pt} 
\sum_{[\mu]}C_{[\mu]}(R_cM_c) 
\prod_{k=1}^{N_D}D^\dagger_{\mu_k}\hspace{6pt}&& 
\label{aqac} 
\end{eqnarray}  
where $A^{\dagger}_{J_q^\sigma M_q^\sigma}(n^\sigma_q)$ and ${\bf 
D}^{\dagger}_{R_cM_c}(N_D)$ are the creation operators of  $n^\sigma_q$  
qp's and $N_D$ $D$-pairs coupled to angular-momenta 
${J_q^\sigma M_q^\sigma}$ and $R_c M_c$, respectively. The 
short-hand notation $[m]$ represents the configuration $\{m_1,m_2, 
\cdots, m_{n^\sigma_q}\}$ of the qp's.  
Similarly, $[\mu]$ is for the $D$-pairs.  
 
\epstblwide{2}{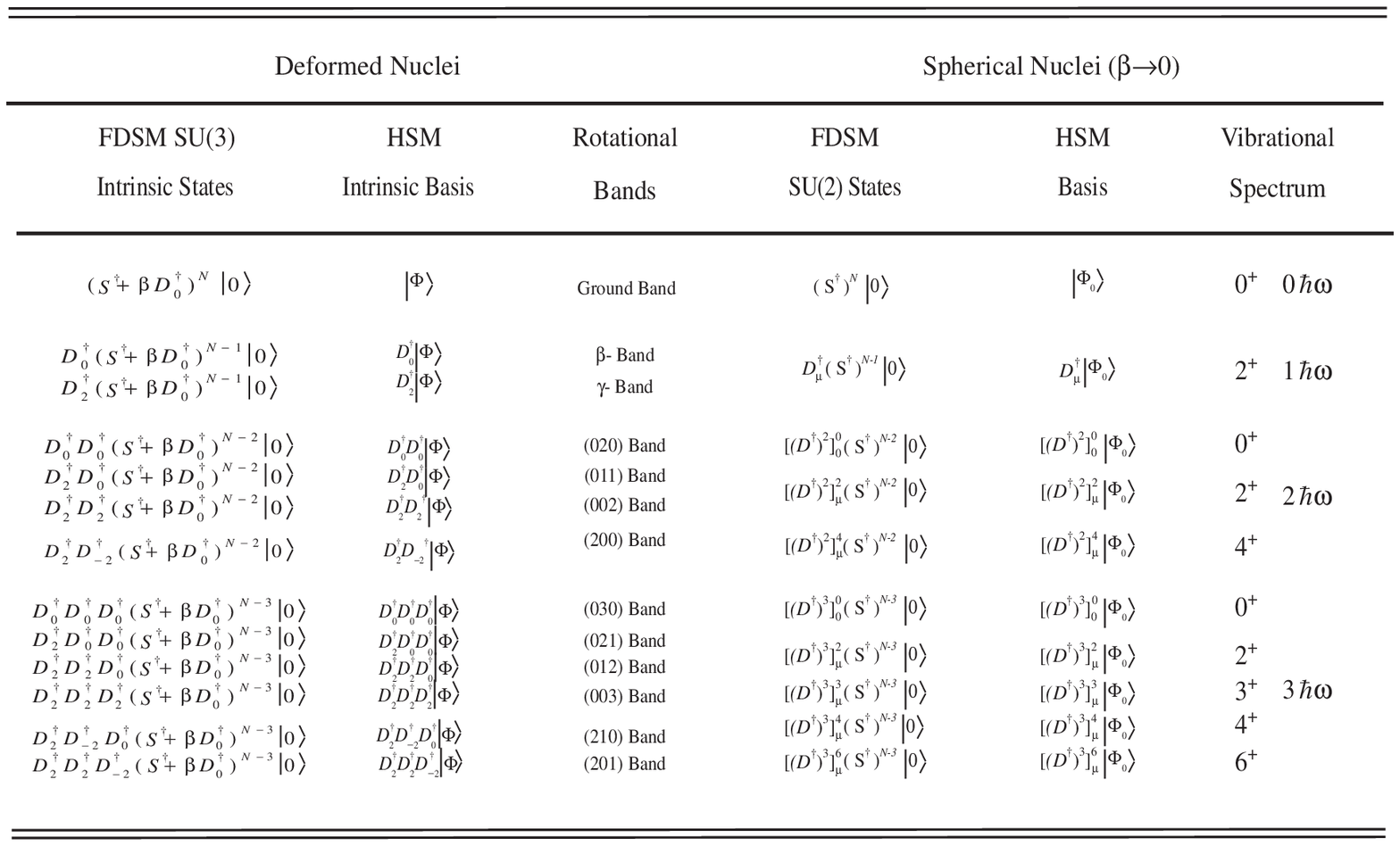}{0pt}{0pt} 
{Comparison of collective basis for deformed and spherical Nuclei. *}  
{* The first column lists the FDSM $SU(3)$ intrinsic states.  
The second column is the HSM collective basis for the 
deformed nuclei, which is obtained by  
replacing $(S^\dagger+\beta D^\dagger_0)^N|0\rangle$ with the BCS 
vacuum $|\Phi\rangle$.  After projection, each intrinsic 
state produces a rotational band labeled as $\beta$, $\gamma$, and 
$(n_\gamma,n_\beta,\kappa/2)$ listed in the    
third column. The FDSM $SU(2)$ states and the HSM collective 
basis for spherical nuclei are listed in the 4th and 5th column,  
respectively.  The last column lists their corresponding spin-parity  
and excitation energies relative to the BCS vacuum. 
Note that the $D$-pairs in the $SU(2)$ states have 
been modified to commute with $S^\dagger$ \cite{fdsm}. } 
 
The amplitudes $C_{[m]}({J_{q}^\sigma M_q^\sigma})$ in Eq.\,(\ref{aqm}) 
can be easily calculated by 
using the standard shell model technique.  As a matter of fact, 
all spherical nuclei lie very close to the doubly closed shell. Therefore,   
only a few qp's from the $j$-orbits around the Fermi surfaces  
need to be considered. 
The amplitudes $C_{[\mu]}(R_cM_c)$ in Eq.\,(\ref{aqac}) 
are difficult to obtain in this way because 
multi-major shells are involved due to the $D$-pair collectivity.  
However, they can be easily evaluated  
through the $d$-boson CFP's (Coefficients of Fractional Parentage)  
if the $D$-pairs are regarded as bosons.   
It should be noted that this is not a boson approximation since  
the $D^\dagger_{\mu}$'s in Eq. (\ref{aqac}) remain 
to be fermionic operators. It is just a symmetry detected truncation to  
select only those states that  
are symmetric with respect to interchanging 
any $D$-pairs. 
 
A comparison between the deformed basis and the spherical basis is shown in  
Table II.  It is interesting to see that the 
$D$-pair excited states, which, in the deformed case, produce the 
$\beta$-$\gamma$ bands after angular-momentum projection,  
degenerate in the spherical case into a spherical vibration-spectrum.  
The spherical vibration-spectrum has 
much less independent states than the rotational spectrum.  It is so  
simply 
because when $\beta\rightarrow 0$, most of  
the states obtained from the angular-momentum projection are not linearly  
independent due to the spherical symmetry nature.   
They are highly over-complete. For instance, the $\beta$- and 
$\gamma$-bands reduce to a $2^+$ state when $\beta\rightarrow 0$.   
This 
is why we should construct the spherical shell-model basis differently.   
It can be seen that even without configuration mixing, this  
collective basis (as shown in Table II)  
can already give 
the essential features of the low-lying collective modes found in the spherical 
nuclei. We therefore feel confident that the HSM basis is going to work well. 
 
Of course, having built a shell-model  
basis is not enough. The central theme that follows is whether it is 
possible to find an efficient truncation scheme within the constructed  
basis so that 
the calculations for heavy nuclei become feasible.  
 
\subsection{Basis Truncation} 
 
To estimate the size  
of the deformed intrinsic basis (\ref{newbasis}), let us denote  
the maximum qp number, the number of selected s.\,p.\ states, and the 
maximum number of the excited $D$-pairs as  
$n^\sigma_{q}$,  $n^\sigma_{s}$ ($\sigma=\nu,\pi$), and $N_m$, respectively.  
The total dimension is the product of the dimension of neutron qp states, 
that of proton qp states, and that of the $D$ pairs 
\begin{equation} 
D_{im}=D^{(q)}_{im}(n^\nu_q,n^\nu_s)\times 
D^{(q)}_{im}(n^\pi_q,n^\pi_s)\times D^{(c)}_{im}(N_m). 
\label{totaldim} 
\end{equation} 
For each of these three terms, it can be shown that  
\begin{equation} 
\begin{array}{l} 
\hspace{.97in}\mbox{\scriptsize  
$n^\sigma_q$}\vspace{-8pt}\\ D^{(q)}_{im}(n^\sigma_q,n^\sigma_s) 
=\sum\hspace{12pt} 
\left( 
\begin{array}{c} 
n^\sigma_s\\ \nu 
\end{array}\right),\hspace{12pt}(\sum_{\nu}\mbox{ is in 
step of 2}) 
\vspace{-9pt} 
\\ 
\hspace{.90in}\mbox{\scriptsize 
$\nu\hspace{-4pt}=\hspace{-2pt}0$ or$1$}\vspace{4pt}\\ 
\hspace{.75in}\mbox{\scriptsize $N_m$} 
\hspace{3pt}\mbox{\scriptsize 
$N_m\hspace{-4pt}-\hspace{-3pt}N^\nu_D$}\vspace{-1.5pt}\\ 
D^{(c)}_{im}(N_m)=\sum\hspace{12pt}\sum\hspace{2pt} 
\left(\hspace{-2pt}\frac{N^{\nu2}_D+4N^\nu_D+3+\delta^\nu}4\hspace{- 
2pt}\right)\hspace{-4pt} 
\left(\hspace{-2pt}\frac{N^{\pi2}_D+4N^\pi_D+3+\delta^\pi}4\hspace{-2pt}\right) 
\vspace{-2pt} 
\\ 
\hspace{.97in}\mbox{\scriptsize 
$\hspace{-20pt}N^\nu_D\hspace{-3pt}=\hspace{-2pt}0$} 
\hspace{24pt}\mbox{\scriptsize 
$\hspace{-20pt}N^\pi_D\hspace{-3pt}=\hspace{-2pt}0$} 
\end{array} 
\label{dim} 
\end{equation} 
where  
$D^{(q)}_{im}(n^\sigma_q,n^\sigma_s)$, with $\sigma=\nu$ or $\pi$, are the 
dimensions of the qp states for neutrons or protons, 
$D^{(c)}_{im}(N_m)$ is the dimension of the intrinsic collective states, 
with $\delta^\sigma=1$ when $N^\sigma_D=$ even and $\delta^\sigma=0$  
when $N^\sigma_D=$ odd.  In 
Eq.\,(\ref{dim}), $D^{(c)}_{im}(N_m)$ is for separated 
neutron and proton vacuum \cite{psmsu3b}. If we treat the neutron and proton 
vacuum as a single coupled BCS vacuum, 
the double-summation in Eq.\,(\ref{dim})  
is reduced to a single one and this can be easily summed up:  
\begin{equation}\hspace*{-6pt} 
D^{(c)}_{im}(N_m)\hspace{-1pt}=\hspace{-1pt}\frac1{24}\hspace{-2pt}\left[\, 
(N_m\hspace{-2pt}+\hspace{-2pt}1)(N_m\hspace{-2pt}+\hspace{-2pt}3)(2N_m\hspace{- 
2pt}+\hspace{-2pt}7)\hspace{-2pt}+\hspace{-2pt}3\delta_{N_m}\,\right 
],  
\label{dimc} 
\end{equation}  
with $\delta_{N_m}=$1 or 0 depending on $N_m=$ even or odd. 
 
According to the experience of the PSM calculations, it is sufficient  
in most cases to  
take $n^\sigma_{q}\le 2$ and $n^\sigma_{s}\le 4$ for both neutrons and 
protons.  One can then obtain that the dimension 
of the qp states $D^{(q)}_{im}(n^\nu_q,n^\nu_s)\times 
D^{(q)}_{im}(n^\pi_q,n^\pi_s)\le 49$.  
If Eq.\,(\ref{dimc}) is used for estimating  
dimension of the collective states,   
$D^{c}_{im}(N_m)$ changes 
from 1, 3, 7, 13 to 161  
when the maximum excited $D$-pair number $N_m$ varies  
from 0, 1, 2, 3, $\cdots$, to 10. 
Thus, the total dimension 
$D_{im}$ ranges from 49, 147, 343, 637, $\cdots$, to 7889.  
 
For well-deformed nuclei, the $D$-pair excitation energy is about 
1 MeV ({\em i.\,e.\,} the band head of the first $\beta$-$\gamma$ band).  
States with more $D$-pairs are excited higher in energy. Furthermore, 
well-deformed nuclei are very close to an $SU(3)$ rotor.  States with different 
$N_D$, which corresponds to different $SU(3)$ irreps, hardly mix with each 
other although the mixing may not be zero since the symmetry is not perfect. 
Thus, for the dominant low-lying collective states the number of the excited 
$D$-pairs, 
$N_D$,  should not be large.  
We expect that taking $N_m=$ 2 to 4 is already good enough to produce 
all the low-lying collective vibrational states.   
Thus, the total dimension $D_{im}$ is of the order of $10^2$ to $10^3$ for 
well-deformed nuclei. 
 
For weakly-deformed nuclei, mixing between different $SU(3)$ irreps must be 
stronger, and therefore,   
it requires a larger $N_m$ to account for such a mixing.   
This can increase the basis dimension drastically. However, 
transition from the deformed to the spherical region behaves like a phase  
transition \cite{Zh99,Ia01}, which happens suddenly within a small  
interval of nucleon number.  
We therefore expect that as we pass through the weakly-deformed region,  
nuclei may quickly enter into the spherical 
region before the $N_m$ number becomes too big (e.\,g.\ $N_m > 10$). 
Thus, calculations for weakly-deformed nuclei, though harder, 
are tractable. 
 
When including the scissors mode in the calculation, we can  
use Eq. (\ref{dim})  
for estimating the basis size. 
The maximum dimension in this case would rise by one to 
two orders of magnitude. Namely, the dimension would become  
49, 294, 1127, 2646, $\cdots$, 2$\times 10^5$  
for $N_m$ = 
0, 1, 2, 3, $\cdots$, 10.  
Moreover, the dimension should be multiplied by a factor 
of $f_{np}$, since, after the neutron and proton states have been built  
with good angular momentum, there are many ways of 
coupling these neutron and proton states  
to a given total angular momentum $I$. It can be 
shown that for a given total angular momentum $I\le 2I_{cut}$, 
\begin{equation} 
f_{np}(I)=\frac I2(4I_{cut}+1-3I)+(I_{cut}+1),  
\label{fnp} 
\end{equation} 
where $I_{cut}$ is the cut-off angular momentum ({\em i.\,e.\,}the  
highest angular momentum that are considered for neutrons and protons).  
This factor is of an order of 
$10^2$ for $I_{cut}\approx 10$. Thus, the total 
dimension $D_{im}$ would be of the order of $10^5$ 
to $10^6$ for $N_m=3$ to 4. However, given the fact that different 
collective modes usually have different energy scale and different symmetries,  
there is still 
room for a further reduction in dimension.  
It is known that the excitation energy of the low-lying 
scissors mode in the well deformed region  
is about 2 MeV higher than the low-lying 
$\beta$-$\gamma$ bands. Furthermore, as shown in Refs. \cite{psmsu3,psmsu3b},  
the scissors mode is physically caused by the relative 
motion of separated n- and p-vacuum (with $N_m=0$),  
not by the $D$-pair excitation. 
Therefore, the interplay between these different collective modes  
should be small 
and may thus be studied separately.  One can thus set 
$N_m=0$ when studying the scissors vibrations. If we include  
qp excitation states up to 4-qp and no $D$-pair excitations in the basis,  
the dimension can be reduced to about 5$\times 10^3$.  
 
For spherical nuclei, the dimension should be estimated differently.  What 
Eqs.\ (\ref{dim}) and (\ref{dimc}) show us is the total $m$-scheme dimension,  
$D_{im}$, in the intrinsic frame. What we are actually interested in is 
the dimension $D_{im}(I)$ at a given 
angular momentum $I$.  
In the deformed case, $D_{im}(I)$ is equal to $D_{im}$ 
because after projection, every intrinsic state can in principle 
contribute one state to an $I$.  
In the spherical case, $D_{im}(I)$  
is smaller than $D_{im}$ since the basis is now 
constructed by angular-momentum coupling in the laboratory frame. 
Eq.\,(\ref{totaldim}) may be utilized to estimate  
the $m$-scheme dimension,  
but the formula for counting the collective basis states 
in the spherical case should be replaced by  
\begin{equation} 
D^{(c)}_{im}(N_m)=\sum_{N_D=0}^{N_m}\left (\begin{array}{c} 
N_D+4\\4 \end{array}\right) . 
\label{dimboson} 
\end{equation} 
This formula is obtained because the collective basis states are 
constructed differently for the spherical case,  
and we have imposed a constraint  
that only those basis states that are symmetric with respect to the interchange of 
$D$-pairs are selected.  According to Eq.\,(\ref{dimboson}),  
for $N_m=0,1,2,3, \cdots,$ 10, the dimension $D^{(c)}_{im}(N_m)$ 
is respectively 1, 6, 21, 56, $\cdots,$ 3003,  
which are considerably larger than the dimensions  
in the deformed case.   
 
An estimation for  
the maximum dimension of 
$D_{im}(I)$, denoted as $D_{im}^{max}$,  
can be made by using the relation  
$D_{im}=\sum_0^{I_{max}}D_{im}(I)(2I+1)$: 
\begin{equation} 
\begin{array}{l} 
D_{im}^{max}=R_I D_{im}\simeq\frac{4D_{im}}{(I_{max}+2)^2}\vspace{4pt}\\ 
I_{max}= I^\nu_{max}+I^\pi_{max}+ I^c_{max} , 
\end{array} 
\label{Dimmax} 
\end{equation} 
where $I^\sigma_{max}$ ($\sigma=\nu, \pi, c$) are the maximum angular momentum 
that the neutron qp's, the proton qp's, and the excited $D$-pairs can reach. 
Thus, knowing the $m$-scheme dimension $D_{im}$ one can estimate  
the maximum dimension $D_{im}^{max}$. The reduction factor 
$R_I=[2/(I_{max}+2)]^2$ makes $D_{im}^{max}$ two to three 
orders of magnitude smaller than the $m$-scheme $D_{im}$.   
 
On the surface it seems that the basis for 
the spherical case is smaller because of the reduction factor $R_I$ in 
Eq.\,(\ref{Dimmax}), it 
is in fact not true.  The problem lies in the fact that  
in the spherical case, the density of  
s.\,p.\ states around the Fermi surfaces is much larger than that 
in the deformed case since each $j$ has a $2j+1$ degeneracy. The number of  
qp states $n^\sigma_s$ must be around 30 instead of 4 adopted in 
the deformed case. This increases the $m$-scheme dimension on the qp sector 
enormously.  Keeping $n^\sigma_{q}\le 2$, the dimension of qp 
basis  will increase from 49 in the deformed case ($n^\sigma_s=4$) to an order  
of $10^5$, leaving almost no room for collective basis.  This is 
another reason why we can not continue to use $m$-scheme basis for the 
spherical case.  Using 
$I$-scheme only $D^{max}_{im}$ should be concerned, which greatly reduces  
the dimension by 
a factor of $R_I$. The total basis dimension varies from 1125, 5879, 17742,  
$4\hspace{-1pt}\times\hspace{-2pt}10^4$, $\cdots$,  
to $1\hspace{-1pt}\times\hspace{-2pt}10^6$  
for $N_m=0,1,2,3,\cdots,$ to 10, when $I^\pi_{max}=I^\nu_{max}=12$ is assumed. 
  
It is known  
that the low-lying spherical vibrational energy 
$\hbar\omega$ is about 0.5 MeV, which is roughly the $D$-pair excitation energy 
in spherical nuclei.  The s.\,p.\ excitations start to have influence at about 
$2\hbar\omega$ and become significant around $3\hbar\omega$, where the 
characteristics of vibrational spectrum are strongly disturbed.  Thus on the 
collective sector, taking $N_m=3$ ($3\hbar\omega$ excitation energy) may be 
a reasonable choice. If the number of qp's is kept  
to be $n^\sigma_q=2$ ($\sigma=\nu, \pi$) to give the maximum 4-qp states,   
the dimension one has to deal with is of the order of 
$10^4$.  
In fact, spherical nuclei lie often near the doubly closed shells 
and have large binding energies. The 4-qp states may already  
be rather high and have 
only little influence to the low-lying states. If we ignore the 4-qp states, 
then the maximum shell-model dimension can be reduced to 488 for 
$N_m=3$.  Even for $N_m=4$ the maximum dimension is only 908. 
Such a basis is clearly manageable. 
 
\subsection{The Effective Interactions} 
 
The HSM Hamiltonian can be generally written in the following form 
\begin{equation} 
\begin{array}{l} 
H=\hspace{6pt}\sum\ \hat H_\sigma+\hat 
H_{\nu\pi},  
\hspace{6pt}\hspace{6pt} 
H_{\nu\pi}=-\sum\ \chi^{(\lambda)}_{_{\mbox{\scriptsize $\nu\pi$}}}\  
\hat Q^\nu_\lambda\cdot\hat Q^\pi_\lambda \vspace{-3pt} \\ 
\hspace{22pt}\mbox{\scriptsize $\sigma=\nu,\pi$}\hspace{1.3in}\mbox{\scriptsize 
$\lambda\ne 1$}\vspace{4pt}\\ 
\hat H_\sigma\hspace{-2pt}=\hat H_0^\sigma\hspace{-2pt}-\hspace{-1pt} 
G^\sigma_{\mbox{\scriptsize M}}\hat 
S^{\sigma\dagger}\hspace{-2pt}\cdot\hspace{-2pt}\hat 
S^\sigma\hspace{-2pt}-\hspace{-1pt} G^\sigma_{\mbox {\scriptsize Q}}  
\hat D^{\sigma\dagger}\hspace{-2pt}\cdot\hspace{-2pt}\hat D^\sigma 
\hspace{-4pt}-\hspace{-1pt}{1\over 2} \sum\  
\chi^{(\lambda)}_{_{\mbox{\scriptsize $\sigma$}}} 
\hat Q^\sigma_\lambda\hspace{-2pt}\cdot\hspace{-2pt}\hat 
Q^\sigma_\lambda\vspace{-3pt} \\ 
\hspace{2.3in}\mbox{\scriptsize $\lambda \ne 1$} 
\end{array} 
\label{Hext} 
\end{equation} 
This rotational invariant  
Hamiltonian is the same as that in the PSM \cite{psm}  
(see also Eq.\,(\ref{H})),  
except that the multipole  
interactions 
are now extended to include not only the quadrupole but also monopole, 
octupole, and hexadecupole terms.  
All these operators are defined in 
the multi-major-shell space: 
 
\begin{eqnarray} 
\hat Q_{\lambda\mu}&=&\sum_{\alpha,\alpha'}\ Q^{(\lambda)}_{\mu\alpha\alpha'}\ 
c^\dagger_\alpha c_{\alpha'}, \nonumber\\  
Q^{(\lambda)}_{\mu\alpha\alpha'}&=&\left\langle\alpha 
|r^\lambda 
Y_{\lambda\mu}(\theta,\varphi)|\alpha'\right\rangle,  
\label{QPext}\\ 
\lambda&=&0,2,3,4.  
\nonumber 
\end{eqnarray} 
The octupole and hexadecupole 
interactions have been employed in their Hamiltonian by Chen and Gao 
\cite{CG01} in dealing with the 
actinide nuclei with projection. 
This type of schematic interactions (\ref{Hext}) 
works for the structure calculations surprisingly well despite its simplicity. 
In fact, it  
has been shown by Dufour and Zuker \cite{Zuker96} that these interactions 
simulate the essence of the most important correlations in nuclei, so 
that even the realistic force has to contain at least these basic components 
implicitly in order to work successfully in the structure 
calculations. Therefore,  
we find no compelling reasons 
for not using these simple interactions. 
Of course, the model is open to adoption of any realistic forces. 
 
The s.\,p.\ energy term $\hat H_0^\sigma$ can be simply taken from the 
Nilsson scheme at zero deformation although other schemes such as the 
Woods-Saxon may also be adopted if there is an advantage.   
The parameterization in the Nilsson scheme has been well established  
for the $\beta$-stable mass regions (see, for example, Ref. \cite{brkm}).  
For the exotic mass regions, the standard Nilsson parameters may not be valid 
and improvement may be necessary \cite{Sn132,Ni56}. 
In this regard, new experimental data that can provide information  
on the single-particle energies in the exotic mass regions are very  
much desired.  The  
Relativistic Mean Field Theory (RMF) \cite{rmf1,rmf2} could also be helpful in 
providing a s.\,p.\ energy scheme for the exotic mass regions where no data are  
available for 
determining a phenomenological mean field. The RMF may have a better 
extrapolation power than other phenomenological mean fields because a single 
set of 
parameters of the RMF  
is able to fit nuclear ground state properties from light to heavy nuclei  
reasonably well.  In practice, it may be a convenient approach  
to use the RMF 
s.\,p.\ energies as a reference to adjust the Nilsson parameters in the HSM.  
 
The parameters 
$\chi^{(\lambda)}_{_{\mbox{\scriptsize $\sigma$}}}$ $(\lambda = 2, 3 ,4)$ 
in Eq. (\ref{Hext}) are determined 
by the self-consistent relation with deformation parameters,   
in the same way as in the PSM \cite{psm,CG01}. 
The monopole-monopole interactions ($\lambda$=$0$) includes three terms:  
$\chi^{(0)}_{_{\mbox{\scriptsize $\sigma$}}}\,n_\sigma$ ($\sigma=\nu,\pi$) 
and $\chi^{(0)}_{_{\mbox{\scriptsize $\nu\pi$}}}\,n_\nu n_\pi$. It has 
been shown that the monopole-monopole interactions are important for 
the nuclear mass calculations, in particular for those nuclei  
lying far from the stability 
line \cite{wumass1,wumass2}. They can be 
viewed as an average way of accounting for the N-Z dependence in s.\,p.\ 
energies. 
However, for a given nucleus, the monopole-monopole interactions contribute a 
constant energy only. Thus,  
if we are not interested in calculating the absolute energies, 
these terms can be ignored.   
The strengths of monopole-pairing, $G^\sigma_{\mbox{\scriptsize M}}$,  
and quadrupole-pairing, $G^\sigma_{\mbox {\scriptsize Q}}$, depend on the 
size of single-particle space. They are inversely    
proportional to the mass number $A$.  
 
For the well-deformed mass regions where calculations have been extensively 
performed 
by the PSM, the interaction strengths proposed for  
the HSM should be similar if the same 
size of single-particle space is employed. 
However, when the HSM is applied to other mass   
regions such as the transitional or 
spherical region, these strengths may need to be 
readjusted. In particular, the self-consistent  
relation used for determining the multipole interaction   
strengths 
will break down when the basis deformation becomes zero.   
The strengths for these cases have to be studied separately.  
After all, Eq.\,(\ref{Hext}) is an effective Hamiltonian 
in nature, and is 
subject to vary when the model is applied to different mass   
regions. Nevertheless, a 
smooth variation in parameterization  
is expected since the model space is sufficiently large.  
 
\subsection{Evaluation of Matrix Elements}  
 
Having had a tractable basis and a reliable effective Hamiltonian,  
the remaining task is to 
diagonalize the Hamiltonian in the basis 
to get the eigen-energies and eigen-functions, and  
then to use the obtained wave functions to calculate the observables.  
To do so, one  
must know how to evaluate the matrix elements in the projected basis.  The 
projection techniques have already been well developed by the PSM  
based on the pioneering work 
of Hara and Iwasaki \cite{HI80}.   
The extensive discussion about the details of the 
projection techniques  
can be found in the PSM review article \cite{psm} and the 
PSM computer code \cite{code}. 
Here we emphasize only on how to deal  
with the new ingredient: the $D$-pairs and the spherical basis.  
 
Let us first discuss the deformed case.  Suppose that a one-body  
tenser operator of 
$\lambda$-rank, $\hat{T}_{\lambda\mu}$, and the eigen-functions  
$|\Psi^I_{M}\rangle$ are given in the laboratory frame: 
\begin{equation} 
\hat{T}_{\lambda\mu}= 
\sum_{\nu,\nu'} 
T^{\lambda}_{\mu \alpha \alpha'} c^\dagger_\alpha c_{\alpha'},\ \, 
|\Psi^I_{M}\rangle=\sum_{qc}F^I_{qc}\hat{P}^{IN}_{MK}|\Phi_{qc}\rangle 
\label{wave} 
\end{equation} 
where $T^{\lambda}_{\mu \alpha \alpha'}\equiv 
\langle\alpha|\hat{T}_{\lambda\mu}|\alpha'\rangle$ is known.  
$|\Phi_{qc}\rangle$ is the intrinsic basis defined in Eq.\,(\ref{newbasis}), and 
the amplitudes 
$F^I_{qc}$ are obtained by solving the eigen-value equation,   
Eq.\,(\ref{sheq}). Although we take the 
tenser operator in Eq. (\ref{wave}) as example, the following discussion applies 
equally well to the pairing 
operators if we change $c^\dagger_\alpha c_{\alpha'}$ in Eq. (\ref{wave}) 
to $c^\dagger_\alpha 
c^\dagger_{\alpha'}$ or $c_\alpha c_{\alpha'}$ and rewrite    
$T^{\lambda}_{\mu\alpha\alpha'} 
\equiv\langle0|\hat{T}_{\lambda\mu}|\alpha\alpha'\rangle$ 
 or $\langle\alpha\alpha'|\hat{T}_{\lambda\mu}|0\rangle$. Sandwiched by the 
wave functions,  the matrix element of 
$\hat{T}_{\lambda\mu}$ in the laboratory frame can be expressed as 
\begin{eqnarray} 
&&\langle \Psi^{I'}_{M'}|\hat{T}_{\lambda\mu} 
|\Psi^I_{M}\rangle=\nonumber\\ 
&&\hspace{24pt}\sum_{q'c'qc}F^{I'}_{q'c'}F^I_{qc} 
\langle \Phi_{q'c'}|\hat{P}^{I'N'}_{K'M'}\hat{T}_{\lambda\mu} 
\hat{P}^{IN}_{MK}|\Phi_{qc}\rangle . \hspace{12pt} 
\label{Tlm} 
\end{eqnarray} 
By applying $[\hat{N}\hat{T}_{\lambda\mu}]=\Delta N 
\hat{T}_{\lambda\mu}$ where $\Delta N = 0$ ($\pm 2$) for the  
electromagnetic multipole (pairing) operator,  
it can be shown that  
\begin{eqnarray} 
&&\hat{P}^{I'N'}_{K'M'}\hat{T}_{\lambda\mu} \hat{P}^{IN}_{MK} 
=\delta^{N'}_{N+\Delta N} 
C^{I'M'}_{IM,\lambda\mu}\nonumber\\ 
&&\hspace{52pt}\times\sum_{\mu'}C^{I'K'}_{IK'-\mu',\lambda\mu'} 
\hat{T}_{\lambda\mu'} \hat{P}^{IN}_{K'-\mu' K}. 
\label{PTP} 
\end{eqnarray}   
Inserting Eq. (\ref{PTP}) into Eq.\,(\ref{Tlm}), we obtain  
\begin{equation} 
\langle 
\Psi^{I'}_{M'}|\hat{T}_{\lambda\mu} |\Psi^I_{M}\rangle= 
C^{I'M'}_{IM,\lambda\mu}\langle 
\Psi^{I'}||\hat{T}_{\lambda}||\Psi^{I}\rangle , 
\label{matrix} 
\end{equation} 
with 
\begin{eqnarray} 
\langle\Psi^{I'}||\hat{T}_{\lambda}||\Psi^{I}\rangle 
&=&\delta^{N'}_{N+\Delta N}\hspace{-9pt} 
\sum_{\mu' 
q'c'qc}\hspace{-8pt}F^{I'}_{q'c'}F^I_{qc}C^{I'K'}_{IK'- 
\mu',\lambda\mu'}\nonumber\\ 
&\times&\langle \Phi_{q'c'}|\hat{T}_{\lambda\mu'} 
\hat{P}^{IN}_{K'-\mu' K}|\Phi_{qc}\rangle 
\label{reduce} 
\end{eqnarray} 
and 
\begin{eqnarray} 
&&\langle\Phi_{q'c'}|\hat{T}_{\lambda\mu'}\hat{P}^{IN}_{K'-\mu'  
K}|\Phi_{qc}\rangle 
=\frac{2I+1}{16\pi^3} 
\int d\phi d\Omega\ e^{iN\phi} \nonumber \\ 
&&\hspace{24pt}\times 
D^{I}_{K'-\mu' K}(\Omega)\  
\langle \Phi_{q'c'}|\hat{T}_{\lambda\mu'}\hat{R}(\Omega,\phi)|\Phi_{qc}\rangle . 
\label{qpmtrix} 
\end{eqnarray} 
Here, $\hat{R}(\Omega,\phi)={R}(\Omega)e^{-i\phi\hat{N}}$ is a rotation 
operator in the 4-dimensional (coordinate plus particle-number) space.  The  
problem is then reduced to evaluate the matrix elements having the general form 
$\langle \Phi_{q'c'}|\hat{T}_{\lambda\nu}\hat{R}(\Omega,\phi)|\Phi_{qc}\rangle$. 
These are the matrix elements for a one-body tenser operator between the 
intrinsic state $\langle \Phi_{q'c'}|$, constructed in (\ref{newbasis}), and  
the rotated one $ \hat{R}(\Omega,\phi)|\Phi_{qc}\rangle$.   
 
In order to evaluate the matrix elements, it is convenient to transform  
the operators $c^\dagger_\alpha$ ($c_\alpha$), contained in 
$\hat{T}_{\lambda\mu}$ and in the $D$-pairs (which are embedded in the basis 
$|\Phi_{qc}\rangle$), into the qp representation $\{a^\dagger, a\}$.  
The transformation matrices (of the Hartree-Fock-Bogoliubov type) are known: 
\begin{eqnarray} 
c^\dagger_\alpha=\sum_\nu 
\left\{U_{\alpha\nu}a^\dagger_\nu-V_{\alpha\nu}\,a_{\bar{\nu}}\right\} 
,&&  
c^\dagger_{\bar{\alpha}} 
=\sum_\nu 
\left\{U_{\alpha\nu}a^\dagger_{\bar{\nu}}+V_{\alpha\nu}\,a_{\nu}\right\} 
\vspace{4pt}\nonumber\\ 
\mbox{with}\hspace{18pt}U_{\alpha\nu}= W_{\alpha\nu}u_\nu 
,&&  
V_{\alpha\nu}=W_{\alpha\nu}v_\nu\ ,  
\label{UV} 
\end{eqnarray} 
where $u$ and $v$ are the occupation amplitudes in the BCS,  
and $W_{\alpha\nu}$'s the 
Nilsson wave functions of the deformed s.\,p.\,level $\nu$. 
 
The evaluation of   
$\langle 
\Phi_{q'c'}|\hat{T}_{\lambda\mu'}\hat{R}(\Omega,\phi)|\Phi_{qc}\rangle$ in 
Eq.\,(\ref{qpmtrix}) is 
equivalent to evaluate the following general  
expectation quantity with respect to the BCS vacuum 
state $|\Phi\rangle$: 
\begin{eqnarray} 
&&\langle 
\Phi_{q'c'}|\hat{T}_{\lambda\mu'}\hat{R}(\Omega,\phi)|\Phi_{qc}\rangle 
=\langle 
\Phi|\cdots D_{\mu2'}D_{\mu1'} \cdots a_{\nu2'}a_{\nu1'} \nonumber \\ 
&&~~\times\ 
\hat{T}_{\lambda\mu'}\hat{R}(\Omega,\phi)\ 
a^\dagger_{\nu1}a^\dagger_{\nu2} \cdots D^\dagger_{\mu1}D^\dagger_{\mu2} 
\cdots |\Phi\rangle . 
\label{qpexp} 
\end{eqnarray} 
After applying the transformation from the spherical to the deformed-BCS 
basis (Eq. (\ref{UV})), The operators appearing in the above expression, such as 
$\hat{T}_{\lambda\mu}$ and $D_\mu$,  
are generally linear combinations of products of  
the qp operators $a^\dagger$ and 
$a$. The problem eventually becomes the one of performing contractions  
for a series  
of qp creation and annihilation operators 
\begin{eqnarray} 
\hspace*{-4pt} 
\langle\Phi|\cdots a_{\mbox{\tiny D2}'}a_{\mbox{\tiny D1}'} 
\cdots a_{\nu2'}a_{\nu1'} \cdots \ 
a^\dagger_{\mbox{\tiny T2}}a_{\mbox{\tiny T1}}\,\hat{R}(\Omega,\phi) 
\nonumber \\ \hspace{4pt}\times\  
a^\dagger_{\nu1}a^\dagger_{\nu2} \cdots a^\dagger_{\mbox{\tiny 
D1}}a^\dagger_{\mbox{\tiny D2}} \cdots |\Phi\rangle . 
\label{contraction} 
\end{eqnarray} 
In performing the contraction calculations,  
a generalized Wick-theorem is used. The techniques  
of carrying out the contractions with the rotation operator 
$\hat{R}$ are available. It has been shown  
\cite{psm,HI80} that the problem can be reduced  
th evaluation the following three basic elements  
\begin{equation} 
A_{\nu\nu'}\equiv\langle a_\nu [\Omega] a^\dagger_{\nu'}\rangle , 
\hspace*{4pt} 
B_{\nu\nu'}\equiv\langle a_\nu a_{\nu'} [\Omega] \rangle , 
\hspace*{4pt} 
C_{\nu\nu'}\equiv\langle [\Omega] a^\dagger_{\nu} a^\dagger_{\nu'}\rangle, 
\label{abc} 
\end{equation} 
 
\noindent 
with $[\Omega] \equiv \hat{R}/ \langle\hat{R} \rangle$, and $\langle\cdots 
\rangle$ is the short-hand notation of vaccum expectation,  
for example, $\langle\hat{O} 
\rangle\equiv\langle\Phi| \hat{O}|\Phi\rangle$.  The contraction can then be 
calculated by the following theorem 
\begin{eqnarray} 
&& \langle a_{n'}\cdots a_{1'}[\Omega] 
 a^\dagger_1\cdots a^\dagger_n\rangle= \nonumber \\ 
&& \hspace{.2in}\sum_{k=m}^{[\frac 
n2]}\sum_p(\pm)(B)^{k-(n-n')/2}(C)^{n-2k}(A)^k \\ 
&& (n+n'=\mbox{even,  } m=max\{0,\frac{n-n'}2\}) \nonumber 
\label{abc2} 
\end{eqnarray} 
where $\sum_p$ is a ``permuted sum" with all possible combinations of pairs 
of indices, and $\pm$ is the permutation parity.  Details for these calculations 
can be found in the PSM 
review article \cite{psm}. 
 
For the spherical case, the matrix elements can be evaluated in the same way. 
The only difference is that there is  
no need for doing angular-momentum projection since 
the basis $|\Phi^{IM}_{qc}\rangle$ defined in Eq.\,(\ref{sbasis}) is 
constructed in the laboratory frame.  
Therefore, in the spherical case,  
the projection operator $\hat{P}^{IN}_{MK}$ should be replaced by  
$\hat{P}^{N}$. As a consequence, Eqs. (\ref{reduce}) and (\ref{qpmtrix}) become  
\begin{eqnarray} 
\langle\Psi^{I'}||\hat{T}_{\lambda}||\Psi^{I}\rangle 
&=&\delta^{N'}_{_{N+\Delta N}}\hspace{-9pt} 
\sum_{\mu' 
q'c'qc}\hspace{-8pt}F^{I'}_{q'c'}F^I_{qc}C^{I'M'}_{IM,\lambda\mu'}\nonumber\\ 
&\times&\langle \Phi^{I'M'}_{q'c'}|\hat{T}_{\lambda\mu'} 
\hat{P}^{N}|\Phi^{IM}_{qc}\rangle 
\label{sreduce} 
\end{eqnarray} 
and 
\begin{eqnarray} 
&&\langle \Phi^{I'M'}_{q'c'}|\hat{T}_{\lambda\mu'} 
\hat{P}^{N}|\Phi^{IM}_{qc}\rangle 
=\frac1{2\pi}\int d\phi d\Omega\ e^{iN\phi}\hspace{18pt} \nonumber\\ 
&&\hspace{.9in}\times\,\langle 
\Phi^{I'M'}_{q'c'}|\hat{T}_{\lambda\mu'}e^{-i\hat{N}\phi}|\Phi^{IM}_{qc} 
\rangle . 
\label{sqpmtrix} 
\end{eqnarray} 
After transforming $\hat{T}_{\lambda\mu'}$ and the $D$-pairs embedded in 
$|\Phi^{I'M'}_{q'c'}\rangle$ and $|\Phi^{IM}_{qc}\rangle$ into the qp basis,  
the evaluation of 
matrix element 
$\langle 
\Phi^{I'M'}_{q'c'}|\hat{T}_{\lambda\mu'}e^{-i\hat{N}\phi}|\Phi^{IM}_{qc}\rangle$ 
is again reduced to the contraction calculations  
like the type of Eq.\,(\ref{contraction}), 
except that in this case, $\hat{R}$ is replaced by $e^{-i\hat{N}\phi}$. 
 
\section{Summary}  
 
In this paper, a multi-shell shell model for heavy nuclei is proposed. 
For performing a shell model diagonalization involving several major  
shells in the model space, we seek  
an efficient truncation scheme.  
The new Heavy Shell Model  
can be viewed as  
an integration of two existing models: the Project Shell Model and the 
Fermion Dynamical Symmetry Model. 
The PSM is an efficient method for the high-spin 
description of rotational states built upon qp-excitations,  
but it is not a practical method for the low-spin collective vibrations. 
In contrast, the FDSM provides a well-defined truncation scheme   
for all known types of 
low-lying collective vibrations, workable from the spherical 
to the well-deformed region, but it is lack of the necessary degrees of freedom 
of single particle excitations. 
The idea proposed in the present  
paper is to combine the advantages of both models.  
To construct the shell model basis, we  
follow the FDSM discovery that the intrinsic collective states can be built 
by applying $D$-pairs onto the BCS vacuum state,  
and employ the PSM qp truncation scheme combined   
with the projection techniques.  
In this sense the model goes beyond the traditional 
one-major-shell shell model, yet the calculation is  
tractable for heavy, and 
even for superheavy nuclei.  
 
Given the past success of the PSM and the FDSM in their own applicable 
regimes, which has been documented in the literature, 
we expect that the new model can work reasonably 
well for heavy nuclear systems  
where traditional shell-model calculations are not feasible.  
This model should be 
capable of describing the low-excitation collective vibrations of all  
known types, including fragmentations due to the quasi-particle mixing. 
It should be 
capable of applying to the high-spin regions where quasi-particle 
alignments play an important role.  
It should also be  
capable of treating weakly-deformed nuclei across the transitional to 
the spherical region.  
We conclude that  
the development of the Heavy Shell Model may open possibilities of 
shell model calculations for heavy nuclei to a much  
wider range of nuclear structure problems.  
 
\section*{Acknowledgments} 
Research at the University of Notre Dame is supported by the NSF under contract 
number PHY-0140324. 
Research at Physics Division of 
National Center for Theoretical Sciences is supported by National Science  
Council in Taiwan. 
 


\begin{thebibliography}{99}
\bibitem{note1} The 
abnormal-parity (intruder) level in each major shell is also taken into account 
in the FDSM, but treated separately. In the context of this paper, 
we shall ignore it for simplicity, since the collective excitations are mainly 
determined by the valence nucleons in the normal-parity levels. The detail 
discussion about the role played by the 
intruder level can be found in the review article of the FDSM \cite{fdsm}.
\bibitem{note2} For convenience, the 
definition of $P_{r\mu}$ here is twice of that in 
Ref. \cite{fdsm}.
\end{thebibliography}
\end{document}